\documentclass[%
reprint,
superscriptaddress,
 amsmath,amssymb,
 aps,
]{revtex4-2}

\usepackage{graphicx}
\usepackage{dcolumn}
\usepackage{bm}
\usepackage{xcolor}

\begin{document}


\title{The Varieties of Schelling Model Experience}

\author{Marlyn Boke}
\author{Timothy Sorochkin}
\affiliation{%
 Department of Physics and Astronomy, University of Waterloo\\
 Institute for Quantum Computing, University of Waterloo
}
\author{Jesse Anttila-Hughes}
\affiliation{
Department of Economics, University of San Francisco
}
\author{Alan O. Jamison}
\email{alanj@uwaterloo.ca}
\affiliation{%
 Department of Physics and Astronomy, University of Waterloo\\
 Institute for Quantum Computing, University of Waterloo
}

\date{\today}

\begin{abstract}

The Schelling model is a prototype for agent-based modeling in social systems. We produce a comprehensive analysis of Schelling model rule variants by classifying the space of macroscopic outcomes using phase diagrams. Among 54 rule variants, only 3 phase diagram classes are found, characterized by the number of phase transitions. This classification scheme is found to be robust to the use of sociological and percolation-inspired measures of segregation. The statistical and dynamic drivers of these transitions are elucidated by analyzing the roles of vision, movement criteria, vacancies, the initial state, and rivalry. Schelling's original step function dictating satisfaction is found to be pathological at high thresholds, producing coordination failures as satisfactory sites become increasingly rare. This comprehensive classification gives new insight into the drivers of transitions in the Schelling model and creates a basis for studying more complex Schelling-like models.
\end{abstract}

\maketitle


\section{\label{sec:Intro}Introduction}

Computational modeling fills a special role in the study of social phenomena, where it is often impossible to design controlled experiments of the sort that form the foundation for understanding in physics. For example, in his 2021 Nobel lecture\cite{card2022design}, Card notes the central need for research design given the impossibility of perfectly observing, controlling, and modeling the complex dynamics that emerge from human behavior. Simulating simplified models of social interactions, particularly using so-called agent-based models (ABMs), allows us to explore possible explanations that may connect us from survey data on individual preferences to statistical results on large-scale social outcomes. The similarities to the core idea of statistical physics---taking microscopic rules and deriving macroscopic behavior from them---have been noted by many authors \cite{Ball2002PhysModelSoc,barthelemy2019StatisticalCities,jusup2022social}.

Advances in the application of statistical physics to social behavior have led to major developments in the study of human cooperation \cite{perc2017statistical} and ``social physics'' more broadly \cite{jusup2022social}. Similar advances in agent-based computational modeling have helped to illuminate the essentially complex nature of economic production \cite{smaldino2016natural,arthur2021foundations} and led to computational heterogeneous agent models becoming the dominant paradigm in monetary macroeconomics \cite{kaplan2018monetary,gabaix2020behavioral}. 

One of the founding ABMs, Schelling's model of segregation\cite{SchelShort, SchelLong, SchelBook}, has been a particularly fruitful point of connection between the physical\cite{bai2022perovskites, PhysAnalogue, PDSchel}, computational \cite{randall2019selforg, randall2022het, Mombaur2016swarm}, and social sciences\cite{tammaru2024socialrev}.  Schelling’s original model highlighted how societal outcomes can collectively fail to match individual preferences in the specific context of housing segregation. Over more than 50 years, the model has been studied using a variety of physical analogues and approaches, and been applied to social problems far beyond its original inspiration \cite{cuevas2020covid}. The broad interest in the model has lead to many new insights, but it also raises questions about the exact nature and goals of the model \cite{Fossett2008social}.

In making analogies to spin systems\cite{Stauffer2008Ising, Nadal2010spin, PDSchel}, binary alloys \cite{PhysAnalogue, Dallasta2007statphys}, active matter physics \cite{zakine2024socioactivematter, seara2025activematterhydro}, and other classic topics of statistical physics, the rules of the original Schelling model have been modified in a variety of ways between studies \cite{absnbs, vision, SchelCityTypeShape, MobilityConstraints}. Investigations from fields outside physics have also introduced variety into the available rule sets followed by the agents. Since the physical study of agent-based social models is still in an early stage of development, it’s important to build formal structures and methods to separate the physics from the unintended consequences of the details of implementing a model.

In this paper we study the differences in macroscopic outcomes that result from minor changes to the rules or initial state used in simulating the Schelling model, which often go unremarked or unexplored. We consider 54 different rule sets inspired by the broad variation found in the rules implemented in the literature. By looking at the global behavior of the model, we classify these rule sets into three categories that demonstrate zero, one, or two phase transitions. 

In section \ref{subsec:Model} we review the workings of the classical Schelling model, being precise about the relevant details of implementation. In section \ref{sec:Methods}, we report our results. These highlight the importance of agent vision and rationality as well as the importance of rival behavior, all of which can be obscured by the details of implementation. In section \ref{sec:Outlook}, we argue that the single phase transition category most cleanly captures the behavior the model is intended to study and discuss implications for comparing and connecting results from the literature. 

\section{\label{subsec:Model}The Model}

\subsection{\label{subsec:OG-Model}`Classic` 2--D Schelling}
In Schelling's original 2--D model \cite{SchelLong}, agents have a group identity and a lower limit for that identity being in the minority within their neighborhood, relocating if the proportion of same-type agents in their neighborhood falls below this homophily threshold. The population is divided into two exhaustive groups, with permanent and recognizable membership for each agent. A square lattice is initialized with such randomly placed agents who are then randomly selected, probed for satisfaction, and relocated to the nearest available satisfactory site if they are found to be dissatisfied.

Schelling's original model has many elements that are readily varied: population density, homophily threshold\cite{PhysAnalogue, granovetter1978threshold}, relocation procedure\cite{PhysAnalogue, MobilityConstraints}, initial configuration, lattice geometry (size and boundaries)\cite{SchelTopology, PhysAnalogue, SchelCityTypeShape}, neighborhood size \cite{vision} and minority status (population ratio)\cite{hatna2010ratio}. 

This paper explores the role of the initial configuration statistics and relocation procedure on the space of segregation outcomes; while maintaining the `classic` agent properties, utility, and the 2D checkerboard location space.

\subsection{\label{subsec:Extended-Model}2--D Schelling Model with Parameter Variations}
\begin{figure*}\label{fig:relocationprocedures}
\includegraphics[width=12.5cm]{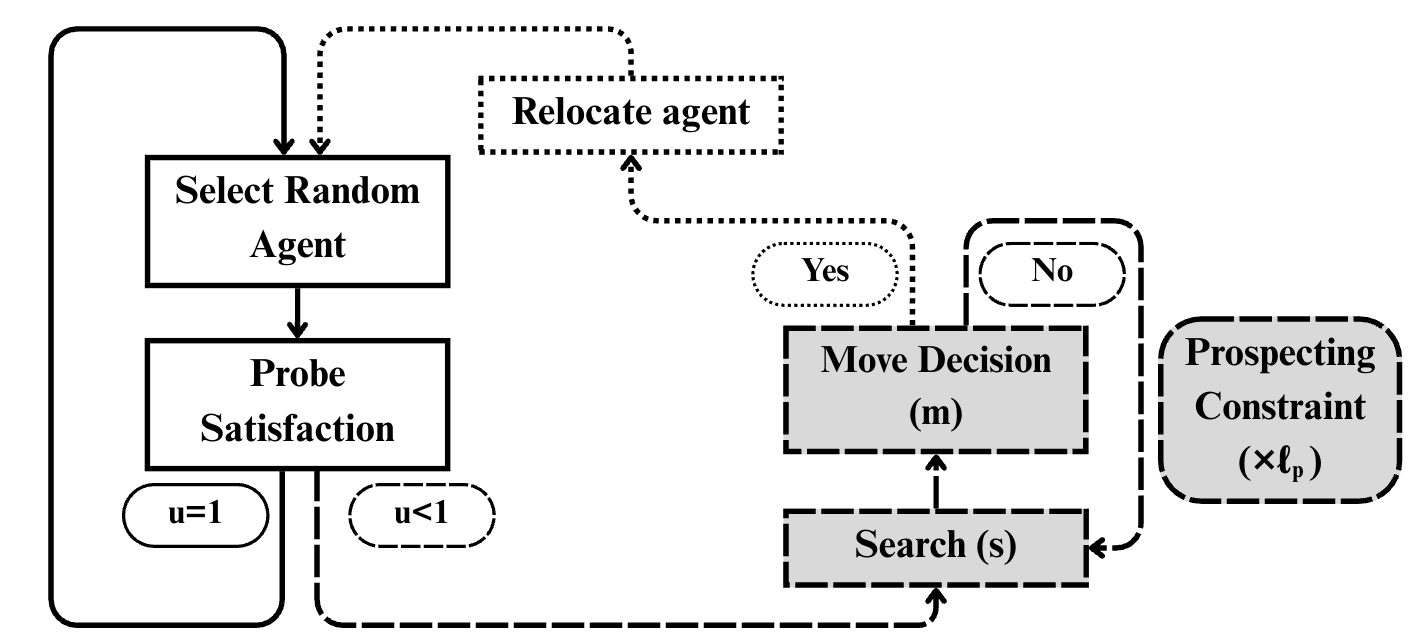}
\caption{General Schelling Relocation Procedure. Agents are randomly probed for satisfaction. Dissatisfied agents evaluate up-to $\ell_p$ sites for places that meet the movement criteria. If the criteria are met, the agent relocates and another agent is randomly selected for probing.}
\end{figure*}
\begin{table*}
\caption{\label{tab:sml indexing} Description of relocation procedure variations and their corresponding [s,m,$\ell_p$] indices}

\begin{ruledtabular}
\begin{tabular}{lcccccr}
\textrm{Rivalry}&
\textrm{Search Procedure}&
\textrm{(s)}&
\textrm{Movement Criteria}&
\textrm{(m)}&
\textrm{Prospecting Constraint}&
\textrm{($\ell_p$)}\\
\colrule
True & Nearest Neighbor & 0 & Move Always & 0 & Single & 1 \\
False & Distance Weighted & 1 & Move Improved & 1 & Finite & $1<\ell_p<\infty$ \\
& Random & 2 & Move Satisfied & 2 & Infinite & $\infty$\\
\end{tabular}
\end{ruledtabular}
\end{table*} 
\subsubsection{\label{subsubsec:City}Location Space and Agents}
A total population of $N$ agents are exhaustively divided into two groups, reds ($N_r$) and blues ($N_b$). These agents are distributed onto a non-periodic square grid where the lattice sites extend from $0$ to $L$ in $x$ and $y$. Agents compete for vacancies, with only one agent allowed per site. 

The expected occupation per site, $\rho$, is thus given by:
\begin{eqnarray} 
\rho = \frac{N}{(L+1)^2} = \frac{N_r + N_b}{(L+1)^2}. 
\end{eqnarray}
\label{eq:occ density}

The global population statistics ($\rho$, $N_{r}$, and $N_{b}$) remain fixed throughout a given simulation. 

\subsubsection{\label{subsubsec:Agents}Agent Properties}
Agents have a location in the 2D lattice $(x,y)$, a type (red or blue), and a homophily threshold (th). 

Each agent's neighborhood is centered on themselves, including all sites within a Euclidean distance $r$ (in lattice units). An agent will count the number of agents of the same ($n_s$) and opposite ($n_o$) type within its neighborhood and determine its homophily quotient $q$ as follows: 
\begin{eqnarray}
q = \frac{n_{s}}{n_s + n_o}.
\label{eq:qsame}
\end{eqnarray} 
Where an agent with $n=n_s+n_o$ neighbors does not consider themselves as a neighbor. If $0\leq q< \text{th}$, the agent is considered dissatisfied, and will relocate. This defines an implicit step function utility describing agent satisfaction in which:
\begin{eqnarray}
u(q) = 
\begin{cases}
    0, & \text{if  $0\le q<$ th}.\\
    1, & \text{if  $q\geq$ th -or- } n=0.
\end{cases}
\label{eq:utility}
\end{eqnarray}

Both the homophily threshold (th) and neighborhood radius ($r$) remain constant throughout the simulation and are uniform across all agents.
\subsubsection{\label{subsubsec:Relocation} Extending Schelling's Model}
The simulations are initialized with a symmetric ($N_r=N_b$) random distribution of N agents on a non-periodic square lattice of size $L=100$; where each agent's neighborhood is of radius $r=10$. 
In each time step, a random agent is selected and probed for satisfaction. For satisfied agents, they are left unperturbed, and a new step is initiated. For dissatisfied agents, the iterative search loop for sites fulfilling the movement criteria is initiated. The search loop terminates if an agent finds a site meeting its movement criteria and thus relocates, or if an agent prospects up to $\ell_p$ sites without success, upon which they give up and are left unperturbed, and a new step is initiated. This process of random probing followed by dissatisfaction driven relocation is repeated until either no relocation takes place, or relocation takes place without altering the overall state of the system. At this point, the simulation is considered to have converged. See Appendix \ref{sec:Conv} for details on the convergence criteria.

Thus, the relocation procedure can be broken into a search algorithm and a movement criterion with a stated upper limit on the number of prospecting attempts, as illustrated in Figure \ref{fig:relocationprocedures}. The rule sets are labeled by s (search algorithm), m (movement criterion), and $\ell_p$ (prospecting loop attempts), respectively, as described in Table \ref{tab:sml indexing}. For a given phase diagram, the relocation procedure ([s,m, $\ell_p$]) and rivalry conditions remain constant throughout all simulations and are uniform across all agents \& lattice sites. The search procedure and movement criteria were found to have the strongest impacts on the phase diagram and as such are the primary variants explored in all the phase diagrams and associated simulations in the main text. Thus, rules are shortened to [s,m] for brevity with single agent site occupation (Rivalry=True) and no upper limit on prospecting time ($\ell_p = \infty$) unless otherwise stated. See \ref{subsec:vacancies} and Appendix \ref{sec:Prospecting} for exploration of multi-agent site occupation (Rivalry=False) and prospecting limits, respectively.

\begin{figure*}
\includegraphics[width=14cm]{ 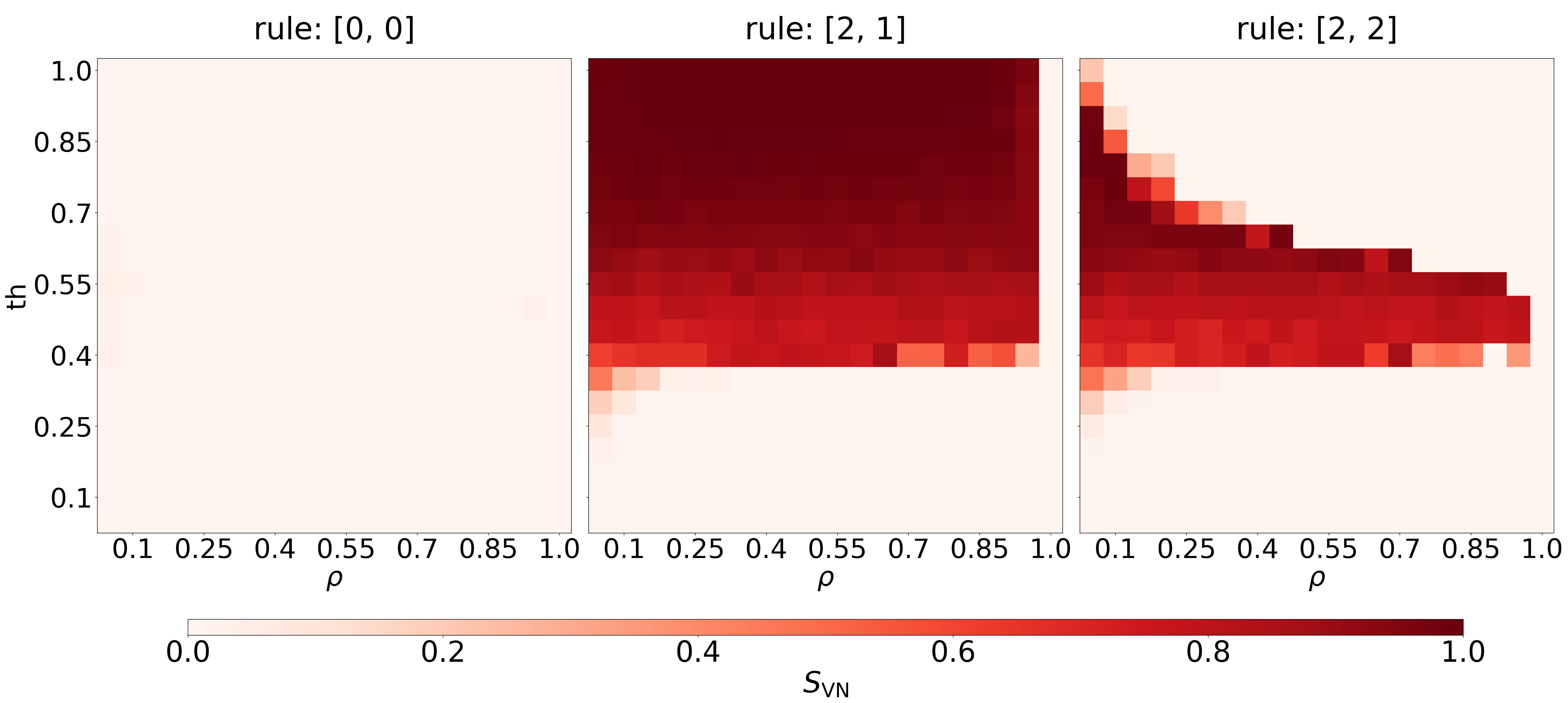}
\caption{\label{fig:3class} Phase diagram classes. We find that three classes of phase diagrams fully classify the space of 54 relocation procedures. Class 0 (left) lacks any transitions away from integrated outcomes, Class 1 (middle) has Schelling's characteristic transition to segregated outcomes at moderately low thresholds, Class 2 (right) has an additional transition away from segregated outcomes despite intense pressure for homophily. All simulations run with Rivalry on, $L=100$ and $r=10$.} 
\end{figure*}

\subsection{\label{subsubsec:PD} Phase Diagrams}
Phase diagrams are used to distinguish regions of distinct qualitative behavior in the space of control parameters. The primary control parameter here is the homophily threshold (th), whose value is shown on the $y$-axis. The secondary control parameter, occupation density ($\rho$), is shown on the $x$-axis unless otherwise noted.

Order parameters describe the global behavior of a given system. The primary order parameter used is a global measure of homogeneity in the lattice, derived from the ensemble average homophily quotient $Q$,
\begin{eqnarray}
Q = \frac{1}{N}\sum_i q_i .
\label{eq:avgqsame}
\end{eqnarray}
For the symmetric population used here, $Q \to 0.5$ for integrated states and $Q \to 1$ for completely segregated states. Thus it can be rescaled as follows:
\begin{eqnarray}
S_{\rm VN} = 2Q - 1,
\label{eq:Svn}
\end{eqnarray}
to calculate the Von-Neumann segregation metric $S_{\rm VN}$\cite{vision} as the primary order parameter. This rescaling of $Q$ ensures that for a symmetric population with a well behaved utility function $ S_{\rm VN} \in [0, 1]$ where:
\begin{eqnarray}
\lim_{Q \to 0.5} S_{\rm VN} = 0 \quad \text{and} \quad \lim_{Q \to 1} S_{\rm VN} = 1.
\label{eq:Svn-limits}
\end{eqnarray}
For alternative measures of segregation, see section \ref{subsec:seg measures}. 

In general, all points on a phase diagram share the same specified relocation procedure (s, m), neighborhood radius, and lattice size \& geometry. These are indicated at the top of the phase diagram. The control parameters are displayed along each labeled axis. The value of the order parameter is represented using a labeled color map.
\section{\label{sec:Methods}Results}
\subsection{\label{subsec:PD}Classification by Phase Diagram}

A comprehensive classification of relocation procedures was achieved by organizing them according to their phase diagram properties. The rule space results in 54 possible relocation procedures that were grouped into 3 distinct phase diagram classes shown in Figure \ref{fig:3class}, each defined by the number of phase transitions. 

The central result of Schelling's work was the transition from integrated outcomes to segregated outcomes at $\text{th} = \vartheta_s$. This transition characterizes class 1, which is defined by this transition alone, typically occurring within the range $0.35 \leq \vartheta_s \leq 0.5$. While agents are satisfied with integrated outcomes within this range of thresholds, agent relocations---driven solely by personal satisfaction and without regard for the impact on neighboring agents---trigger a cascade of subsequent dissatisfaction-driven moves. This cascade, referred to as Schelling's avalanche henceforth, tips the outcome toward segregation, since it is the dynamically stable outcome. The sensitivity of Schelling's avalanche to agent vision is discussed in Section \ref{subsec:vision}. Of the 54 generated phase diagrams 14 ($26\%$) corresponded to class 1. 

Class 2 is characterized by $\vartheta_s$ and an additional transition away from segregated outcomes at $\text{th}=\vartheta_m$. This transition typically occurs within the range $0.5 \leq \vartheta_m \leq 0.8$ and is driven by coordination failures discussed in Section \ref{subsec:rationality}. Of the 54 generated phase diagrams, 20 ($37\%$) corresponded to class 2.

The final class---class 0---lacks any transitions away from integrated outcomes, as $\vartheta_s \simeq \vartheta_m$. This is revealed by examination of other order parameters such as dynamic activity. Class 0 is found to often be an edge case of the Nearest Neighbor searches phase diagrams in which $\vartheta_s, \vartheta_m \sim 0.5$ due to constrained prospecting vision paired with either a blind move rule or constrained prospecting time. This results in a phase diagram characterized by integration at all thresholds, lacking any obvious phase transitions in segregation. Of the 54 generated phase diagrams, 10 ($18.5\%$) corresponded to class 0. There are an additional 10 ($18.5\%$) phase diagrams belonging to distance weighted search algorithms (see Figures \ref{fig:DW} and \ref{fig:class0 exp}) whose categorization varies from class 0 to class 2 depending on the search radius used. 

In the following sections, the statistical and dynamical drivers of these transitions are elucidated. In section \ref{subsec:vision}, the role of vision on the lower transition at $\vartheta_s$ in classes 1 and 2 is explored. The role of vision is then used to contextualize both Schelling's original result in section \ref{subsubsec:Schell-results} and the collapse of class 2 into class 0 in section \ref{subsec:class0}. In sections \ref{subsec:rationality} and \ref{subsec:vacancies}, the role of the decision rule and vacancies on the upper transition at $\vartheta_m$ in class 2 are explored. Lastly, in section \ref{subsec:seg measures}, segregation measures inspired by percolation theory, the study of clustering, are explored to investigate whether alternative measures reveal unobserved classes of behavior. 
\subsection{\label{subsec:vision}Role of Vision}
Agents possess two types of vision, the ``prospecting vision'' used to decide \textit{where} to move and the ``neighborhood radius'' $r$ used to decide \textit{when} to move \cite{vision}.
\subsubsection{\label{subsubsec:prospecting-vision}Prospecting Vision}
When searching, agents randomly select a prospect site within some search radius. Schelling's early model employed Nearest-Neighbor (NN) search (s=0), prioritizing proximity by prospecting the closest sites before searching farther away. Conversely, much of modern Schelling research employs an entirely random search (s=2) with no preference for proximity. The prospecting vision was found to affect the location of the lower transition $\vartheta_s$ due to the relationship between agent mobility and the Schelling avalanche.

\begin{figure}
\includegraphics[width=8.7cm]{ 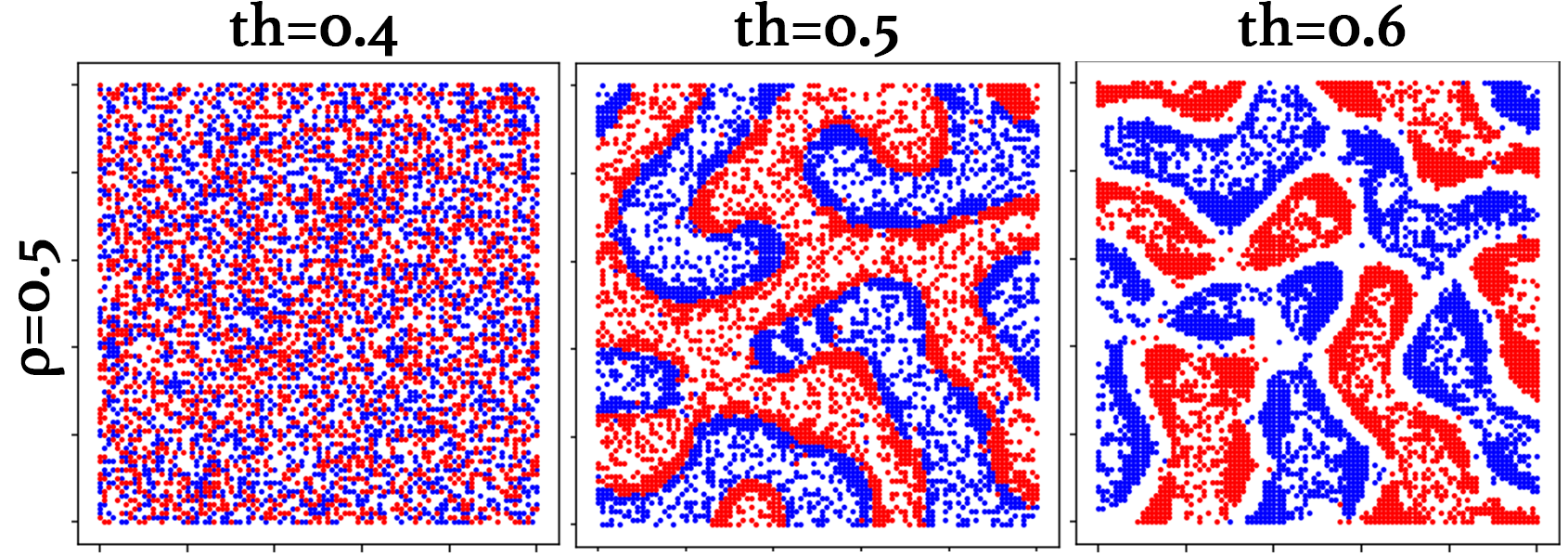}
\caption{\label{fig:NN tiles} NN Search Segregation patterns. With nearest neighbor search at $\text{th} > 0.4$, we find final states characterized by finger-like clusters with large surface areas and internalized vacancies at moderate thresholds. As homophily demands increase, vacancies move to act as buffers between opposite type clusters.} 
\end{figure}

\begin{figure}
\includegraphics[width=8.7cm]{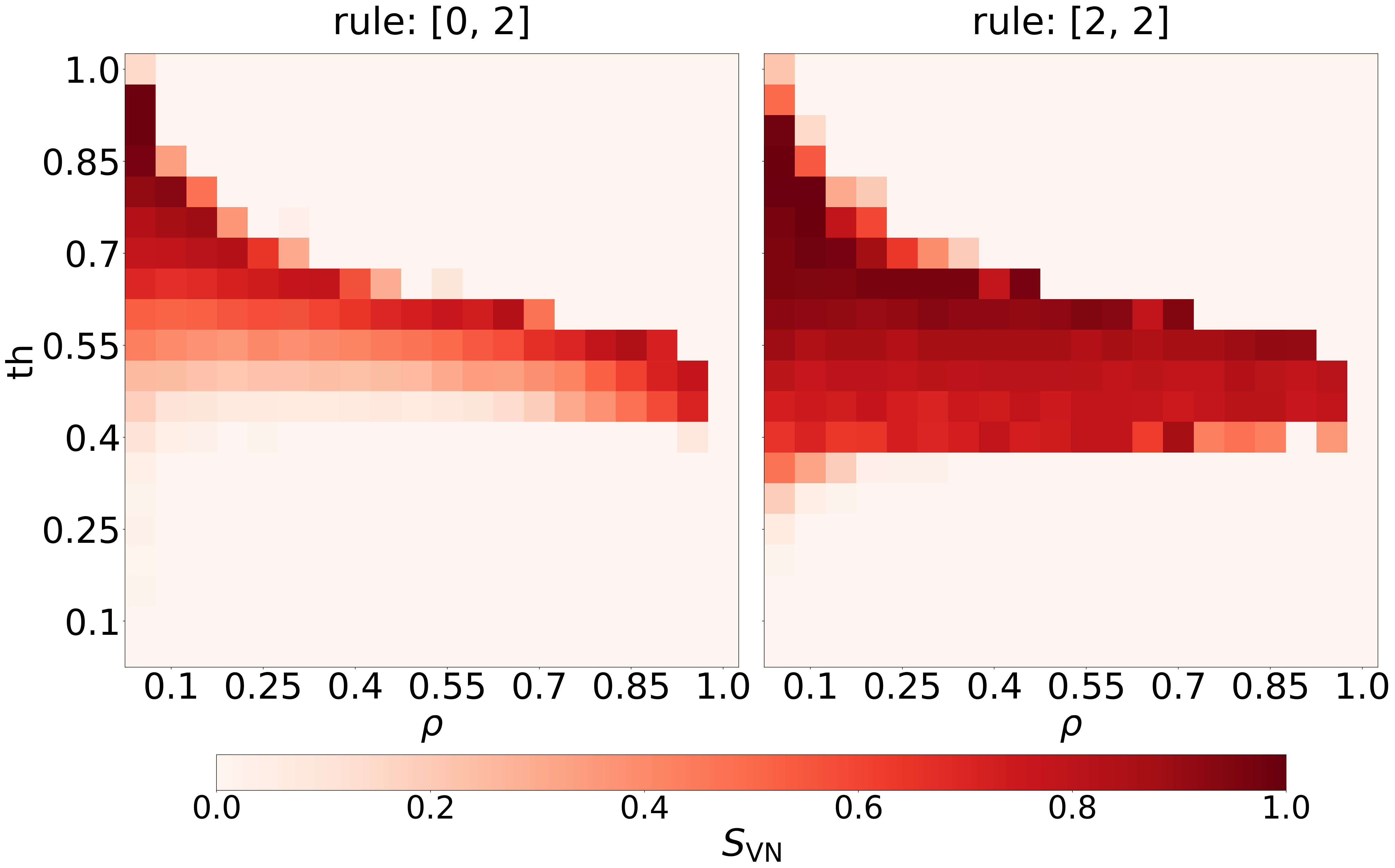}
\caption{\label{fig:vision} Satisfaction driven Class 2 phase diagrams. Results are shown for satisfaction seeking agents ($m=2$) under NN search (left: $s=0$) and random search (right: $s=2$) procedures. NN search spatially constrains avalanche dynamics yielding a larger $\vartheta_s$ in comparison to random search. As the lattice fills, agents must search further out, causing the two search algorithms to converge at high density. All simulations run with Rivalry on, $L=100$ and $r=10$.} 
\end{figure}

In the case of nearest-neighbor (NN) search, the avalanche triggered by agent relocation is localized to nearby neighborhoods. This limited mobility behaves like a slow diffusion rate \cite{PhysAnalogue}, where agents only move the minimum possible distance necessary to achieve movement criteria. These agents relocate to sites in close proximity, which often share a large area of the agent's original neighborhood and only enable small changes in $q$ due to spatial correlations. As a result, the final state remains closer to its initial configuration and only a weaker, more localized form of segregation is attained at $\text{th} > \vartheta_s$. This is evident in the finger-like clusters present at moderate thresholds in Figure \ref{fig:NN tiles}. With a larger surface area of contact between the two groups, a lower Von-Neumann measure of segregation is achieved at $\text{th}=0.45$ in Figure \ref{fig:vision}. Ultimately, constraining mobility spatially contains the avalanche as a dissatisfied agent only destabilizes neighborhoods in close proximity; this results in slightly higher transition threshold, with $\vartheta_s \simeq 0.45$.

\begin{figure}
\includegraphics[width=8.7cm]{ 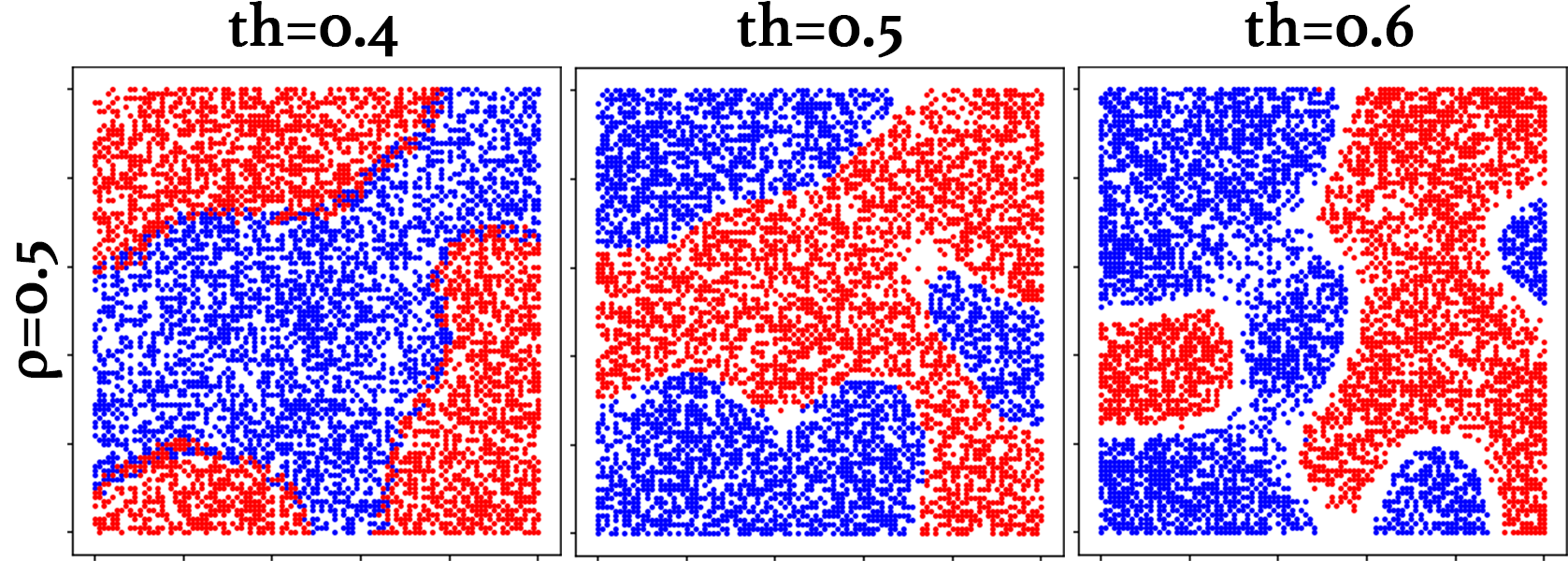}
\caption{\label{fig:random tiles} Random Search Segregation Patterns. With random search, consolidated clusters with smaller surface areas and uniformly distributed vacancies are formed at moderate thresholds (th $> 0.35$). As homophily demands increase, vacancies move to act as buffers between opposite type clusters. } 
\end{figure}

In contrast, when prospecting is unconstrained---as in the case of random search---the destabilizing effect of agent relocations is lattice-wide, as agents are able to relocate freely across the system. This unrestricted mobility allows for super-diffusive motion, allowing agents to easily overshoot their goals. As a result, the final state strays further away from the initial configuration, producing a much more consolidated form of segregation (see Figure \ref{fig:random tiles}) compared to the structures observed under NN search. Consequently, this untamed avalanche dynamic triggers segregation at a lower threshold, with $\vartheta_s \simeq 0.35$.

Prospecting vision also influences the role of vacancies in the range $\vartheta_s < \text{th} < 0.5$. Under nearest-neighbor search, vacancies play a crucial role in enabling diffusion and are thus actively driven into the core of clusters as agents move away from dissatisfactory locations. This results in clusters with the highest agent density on the surface.  In contrast, with random search, movement is sufficiently rapid at moderate vacancy densities to make this mechanism unnecessary; thus, vacancies remain uniformly distributed. As homophily demand increases, the role of vacancies as a buffer between opposite-type groups becomes increasingly important at higher thresholds under both search procedures. Finally, at higher occupation densities and thresholds, the behavior of the two models fully converge as few sites meeting movement criteria are available nearby and agents are forced to move further away, resulting in greater diffusion out of necessity.
\begin{figure}
\includegraphics[width=7.5cm]{ 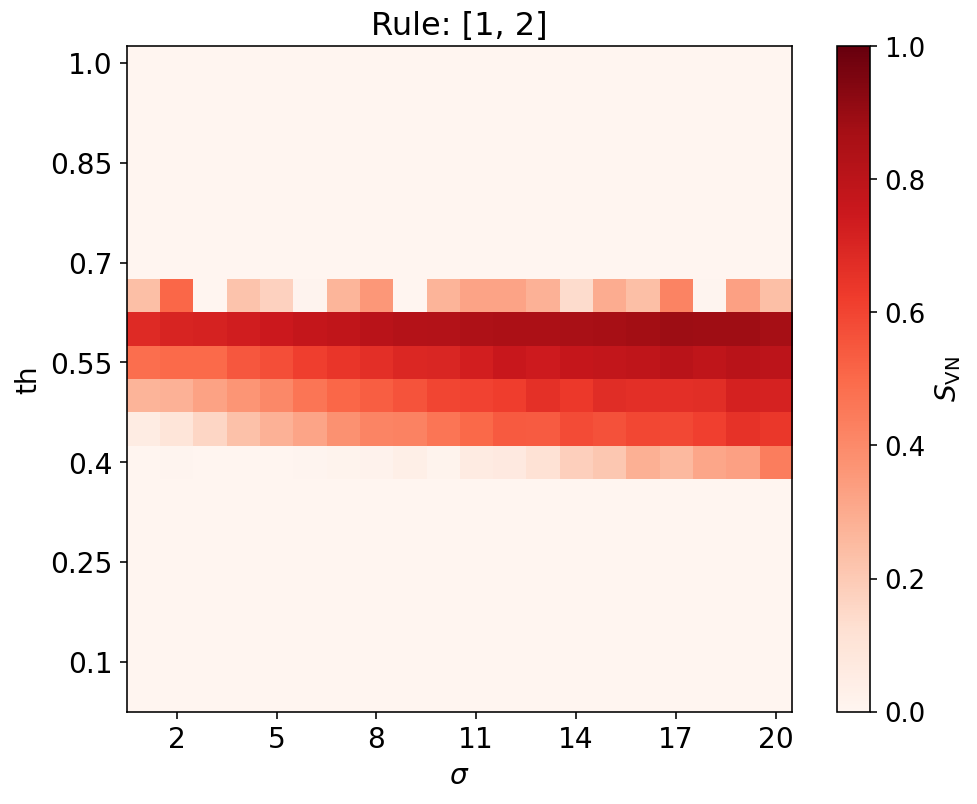}
\caption{\label{fig:DW} Effect of varying agent prospecting vision. Results shown for satisfaction driven agents ($m=2$) employing a distance weighted search ($s=1$). Here, the search radius ($\sigma$) can replace $\rho$ as a secondary control parameter. As the search radius expands, $\vartheta_s$ transitions smoothly from its value under NN search to its value under random search. All simulations run with Rivalry on, $L=100$, $r=10$ and $\rho=0.5$.}
\end{figure}

Additionally, the lower transition in Figure \ref{fig:vision} is sensitive to $\rho$ due to the secondary role of vacancies in regulating neighborhood occupation (see section \ref{subsec:vacancies}). As neighborhood occupation ($n$) increases, fewer agents in the initial state are dissatisfied (see Figure \ref{fig:qsame histogram}), and these agents' relocations have less of a perturbating effect on the agents in the immigrant or emigrant neighborhoods. Thus, the seeds of dissatisfaction in the initial state will tip fewer neighborhoods into becoming unsatisfactory. This results in $\vartheta_s$ increasing to $0.4$ under random search. For NN search, as $n$ increases fewer sites are available in close proximity and thus NN search and random search converge as agents are forced to move further away. This brings $\vartheta_s$ down from $0.45 \to 0.4$ matching that of random search.


To connect these two extremes of prospecting vision, a distance-weighted search (s=1) is implemented. Under this search procedure, the likelihood of selecting a site for prospecting follows a Gaussian probability function with variance $\sigma^2$ such that sites within a distance of $\sigma$ are more likely to be selected.
Notably, no distinct intermediate regime emerges between the NN search ($\lim \sigma \to 1$) and random search regimes ($\lim \sigma \to \sqrt{2}L$) in Figure \ref{fig:DW}. Instead, $\vartheta_s$ decreases smoothly, connecting rules [0,2] and [2,2] at $\rho=0.5$. As discussed in Appendix \ref{sec:Measures}, different segregation metrics show even more clearly how smoothly the distance-weighted search interpolates between the two other rules.




\subsubsection{\label{subsubsec:radius}Neighborhood radius}
\begin{figure}
\includegraphics[width=8.5cm]{ 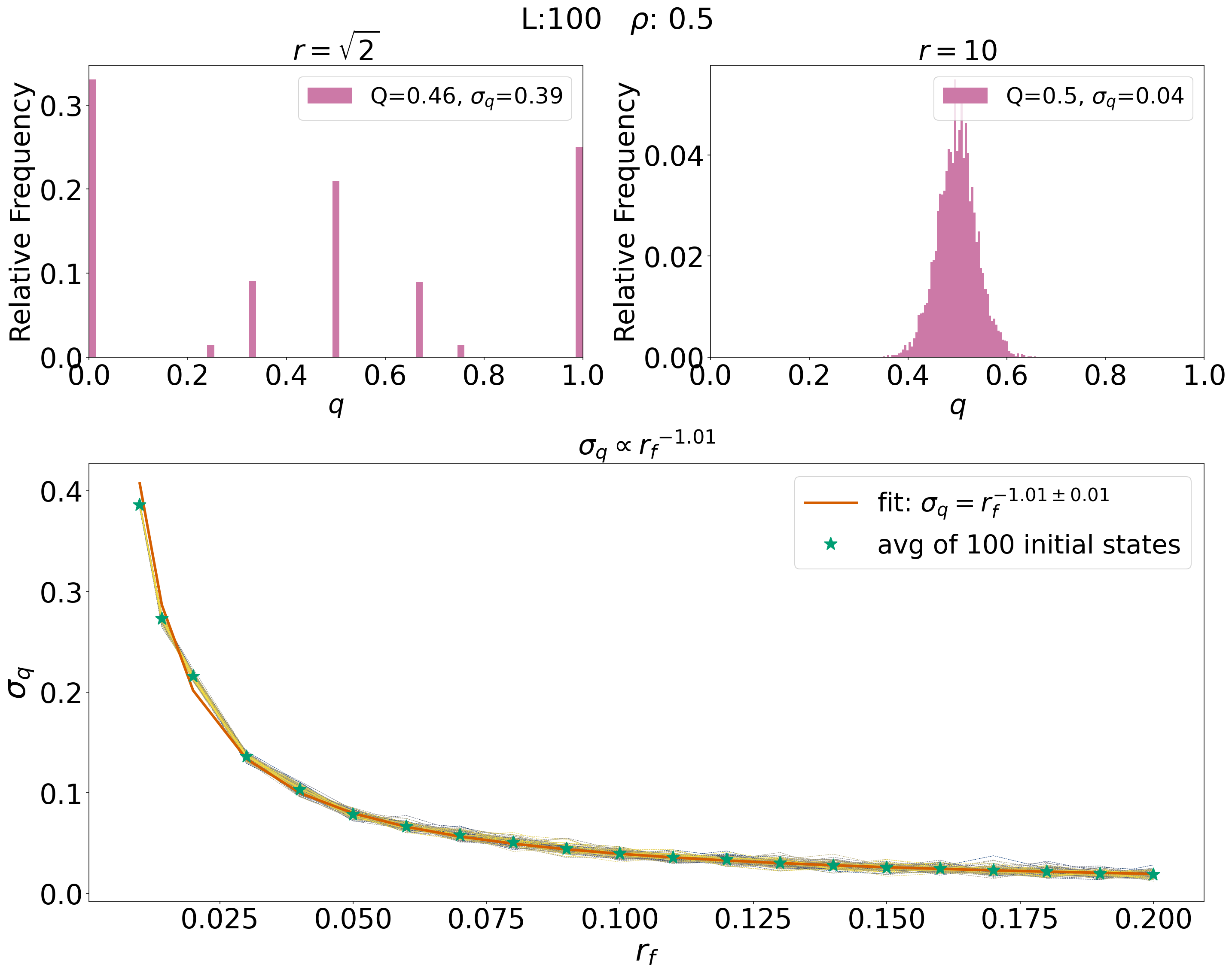}
\caption{\label{fig:qsame histogram} Distribution of initial state homophily statistics ($q$). Distribution of agents homophily quotients in a randomly distributed initial state ($L=100, \rho=0.5$) for r=$\sqrt{2}$ (top left) and r=$10$ (top right). Empirical examination of 100 initial states found the standard deviation of the homophily distribution to be inversely proportional to neighborhood radius (bottom panel) with a $\ln \sigma_q$ vs $\ln r_f$ fit yielding $\sigma_q = r_f^{-1.01 \pm 0.01}$. Distributions were generated on a periodic lattice.} 
\end{figure}

\begin{figure}
\includegraphics[width=8cm]{ 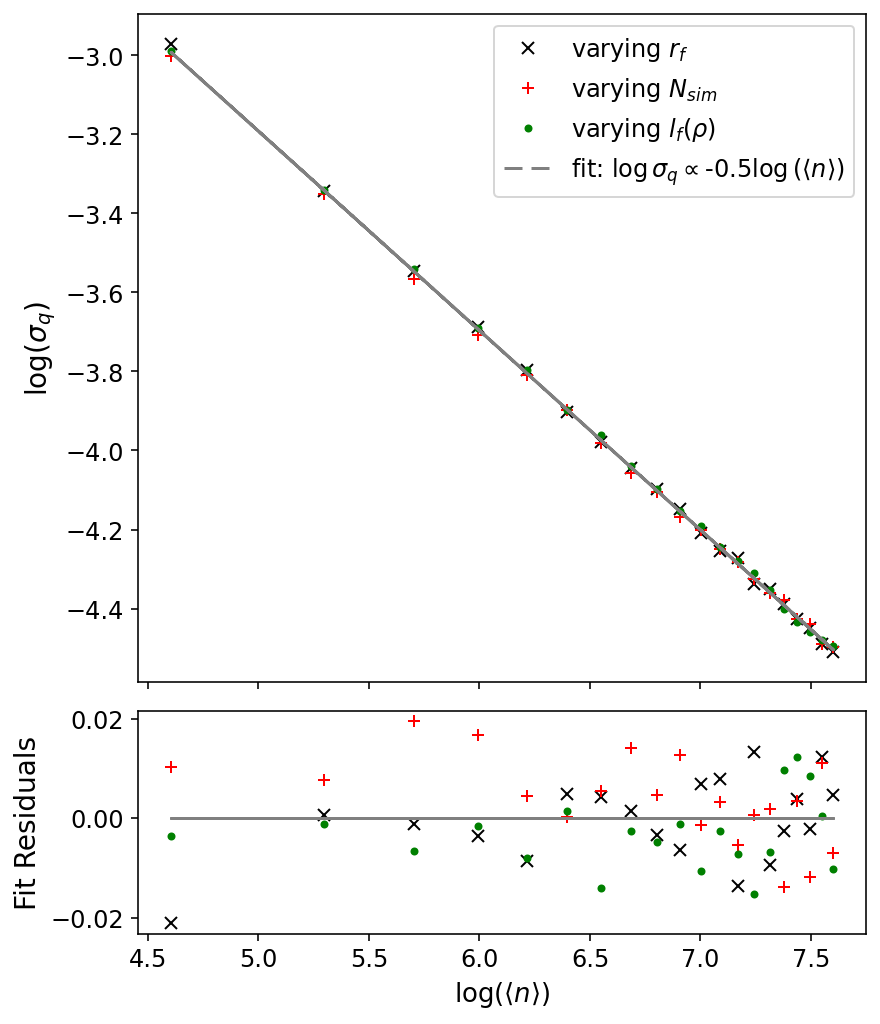}
\caption{\label{fig:fluctuations-all} Neighborhood occupation density controls initial state statistics. The various control parameters for neighborhood occupation are varied independently of one another: neighborhood $r_f$ (fixed $N_{sim}, L$), $N_{sim}$ (fixed $r_{f}, L$), and lattice size (fixed $r, N_{sim}$). Fit Residuals for each dataset are shown in the bottom panel. All distributions were generated on a periodic lattice.}
\end{figure}
\begin{figure}
\includegraphics[width=8.7cm]{ 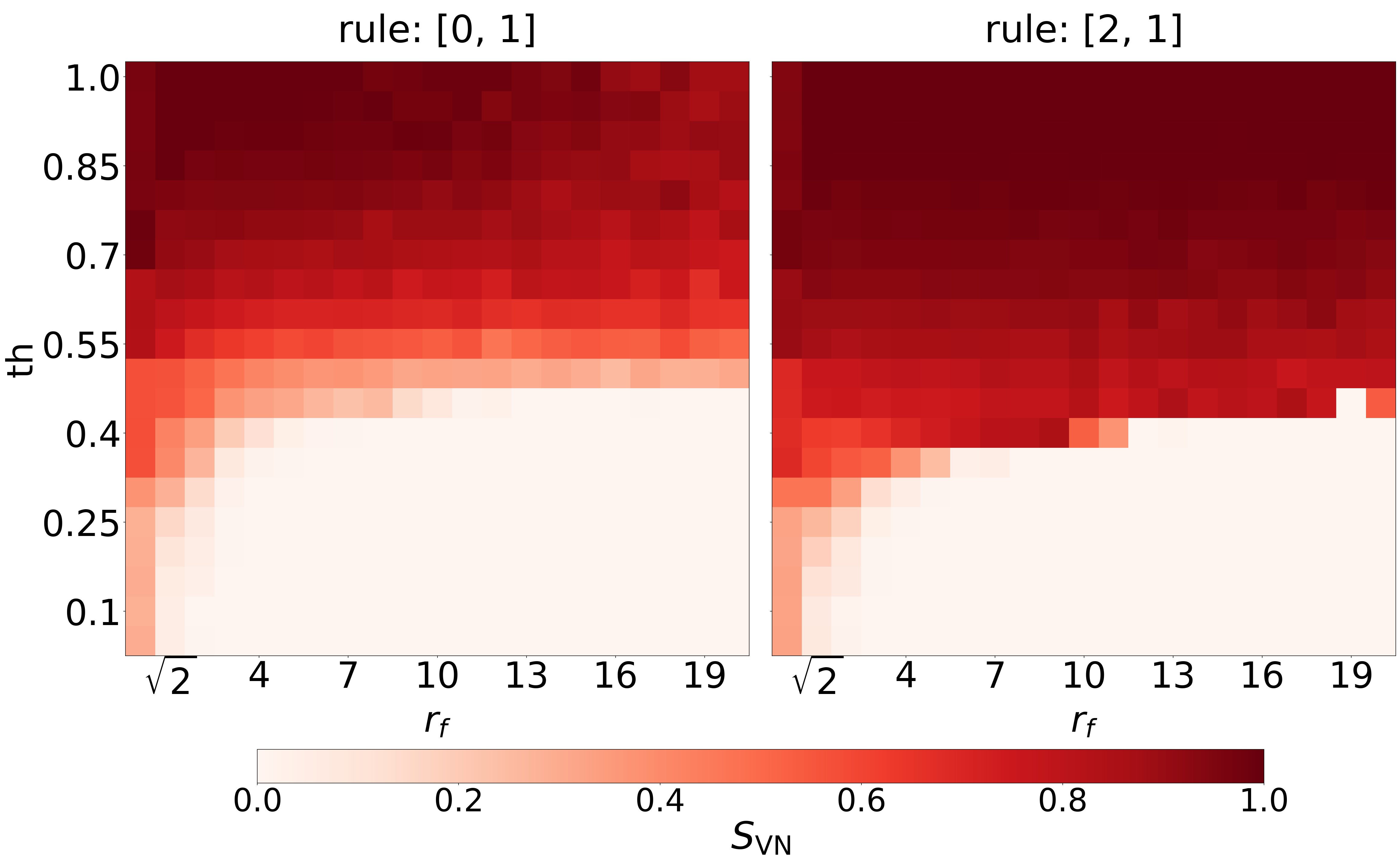}
\caption{\label{fig:rlatPD} Effect of an agent's neighborhood radius on lower transition. Neighborhood radius ($r$) can replace $\rho$ as a secondary control parameter. Results are shown for NN search (left: $s=0$) and random search (right: $s=2$) procedures. As $r$ increases, there is less variance in the initial state homophily quotients, providing fewer seeds and damping Schelling's avalanche to yield a larger $\vartheta_s$. All simulations run with Rivalry on, $L=100$ and $\rho=0.7$.} 
\end{figure}

To study the role of the neighborhood radius, a phase diagram was generated that varied $r$ from 1 to 20 as illustrated in Figure \ref{fig:rlatPD}. A smaller $r$ results in a more discretized range of possible $q$, with larger spacing between values. 

Additionally, a smaller $r$, at fixed $L$ and $\rho$, decreases the average number of agents in a neighborhood $\langle n \rangle$.This results in larger fluctuations in the initial state as:
\begin{eqnarray} 
\sigma_q ^ 2 \propto \frac{1}{\langle n \rangle}.
\label{eq:fluctuations}
\end{eqnarray}
Here the expected number of agents in a neighborhood $\langle n \rangle$ is:
\begin{eqnarray} 
\langle n \rangle = \rho \pi r^2 \simeq  N_{sim}\pi r_f^2
\label{eq:nbhood density}
\end{eqnarray}

To isolate the role of the neighborhood size from the lattice size, $\sigma_q$ was plotted as a function of the relative neighborhood size $r_f = r/L$. Note that since $r^2 \propto L^2$ and $\rho \propto L^{-2}$, any lattice size dependency in $\sigma_q$ is canceled out due to these parametrization choices. Equation (\ref{eq:fluctuations}) is verified empirically in Figures \ref{fig:qsame histogram} \& \ref{fig:fluctuations-all}, and derived in Appendix \ref{sec:fluctuations}. This confirms the need for systematic analysis of the model's sensitivity to initial conditions \cite{sensitivity2018}. Further, it agrees with the findings of Fossett et al \cite{SchelCityTypeShape}, in which neighborhood occupation affected segregation outcomes, but lattice size (independent of $r$) had no effects.

Thus, smaller $r_f$ yields larger $\sigma_q$, increasing the likelihood of finding unsatisfied agents in the initial state. Simultaneously, the effect of these agents relocations are amplified, as neighborhoods with small $n$ are perturbed more significantly by the addition or loss of an agent. As a result, at smaller $r$ the transition towards segregation occurs at smaller $\vartheta_s$ as there are more seeds for Schelling's avalanche, and the destabilizing effect of the avalanche is amplified.

\subsection{\label{subsubsec:Schell-results}Schelling's Original Phase Diagram}
\begin{figure}
\includegraphics[width=7cm]{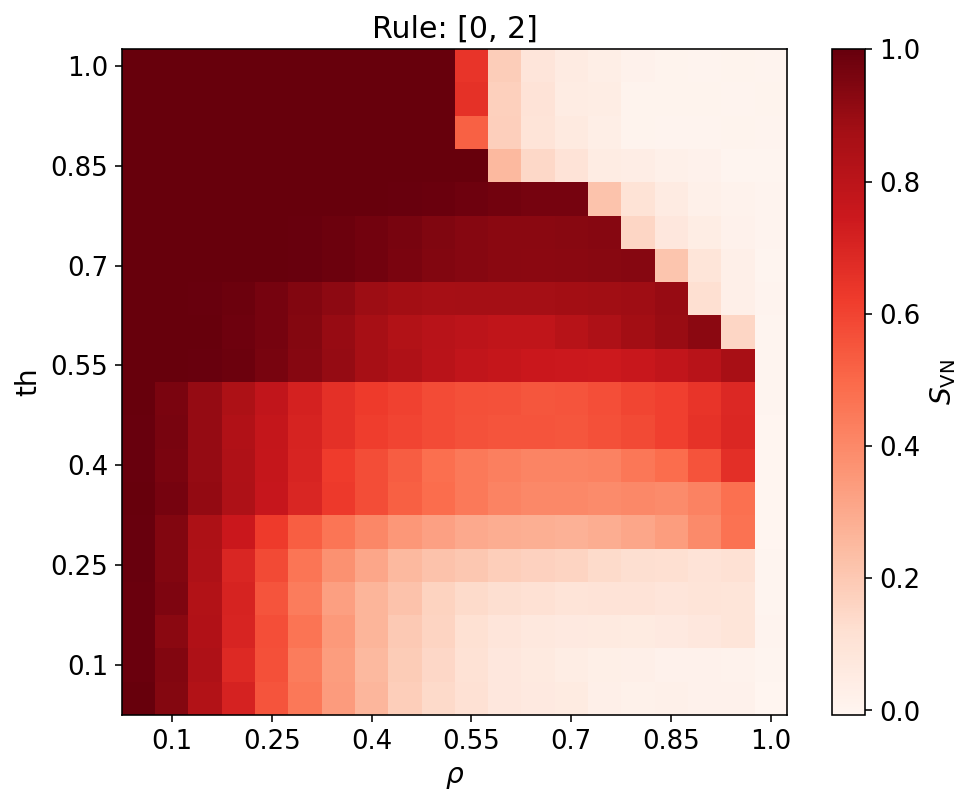}
\caption{\label{fig:SchelOG} Schelling's original phase diagram. The lower transition occurs at $\vartheta_s \simeq 0.3$. While Schelling's use of NN search would tend to constrain the avalanche, the small neighborhood size ($r = \sqrt{2}$) minimizes correlations in $q$ between subsequent prospects, provides more seeds, and amplifies the destabilizing effect of a seed's relocation. All simulations run with Rivalry on, $L=100$ and $r=\sqrt{2}$.} 
\end{figure}

Schelling's relocation rule was nearest neighbor search for available satisfactory sites, corresponding to rule [0, 2], performed with a smaller neighborhood radius ($r_M=\sqrt{2}$) known as the Moore neighborhood. For the 54 rule variants performed with $r=10$, at moderate to low occupation densities when prospects can be found within $d<2r$ the spatial correlations between the current site and prospect sites' neighborhoods yield only small changes in $q$ as agents relocate, resulting in a constrained avalanche and a delayed transition at $\vartheta_s \simeq 0.5$. However, for Schelling's Moore radius, at moderate occupation densities ($\rho > 0.15$), few to no vacant prospects within $d\leq2r_M$ can be found. Viable prospects ($d>2r_M$) consequently have no spatial correlation with the original neighborhood, yielding $\vartheta_s \simeq 0.3$ once more---the puzzling result that started all of this work (see Figure \ref{fig:SchelOG}). At low occupation densities, when many vacancies can be found within $d<2r_M$ of any agent, very few agents are within $d<2r_M$ of one another. These isolated agents dominate the ensemble average homophily quotient, yielding a segregated measure at all thresholds.


\subsection{\label{subsec:rationality}Role of the Decision Rule}

Schelling's early analysis focused on utility-maximizing agents that only move to satisfactory sites ($m=2$). Alternatively, a model with agents that ignore satisfaction goals and move as long as the site is available (i.e., vacant) ($m=0$) has been explored here. Both of these regimes of movement criteria result in integration at $\text{th}>\vartheta_m$. Despite the strong homophily preference, agents are unable to achieve satisfactory outcomes due to coordination problems arising in the search for satisfactory locations. The nature of these coordination problems is elaborated in sections \ref{subsubsec:restless} and \ref{subsubsec:petrified}.

These coordination problems result from an implicit utility that only enables agents to improve by finding satisfactory locations, resulting in pathological outcomes at high thresholds. To test this, the implicit utility function was relaxed as follows:

\begin{eqnarray}
u(q) = 
\begin{cases}
    \frac{q}{\text{th}}, & \text{if  $0\le q<$ th},\\
    1, & \text{if $q\geq$ th -or- } n=0 ,
\end{cases}
\label{eq:utility2}
\end{eqnarray} 

to allow sites with larger $q$ to be viable for relocation. This enables utility-maximizing agents with intense homophily goals to ``walk toward satisfaction'' by exploiting fluctuations in the initial state to find neighborhoods with higher but not yet satisfactory $q$. This movement, in turn, reinforces and expands these favorable areas, ultimately enabling agents to achieve satisfaction.

\begin{figure}
\includegraphics[width=8.7cm]{ 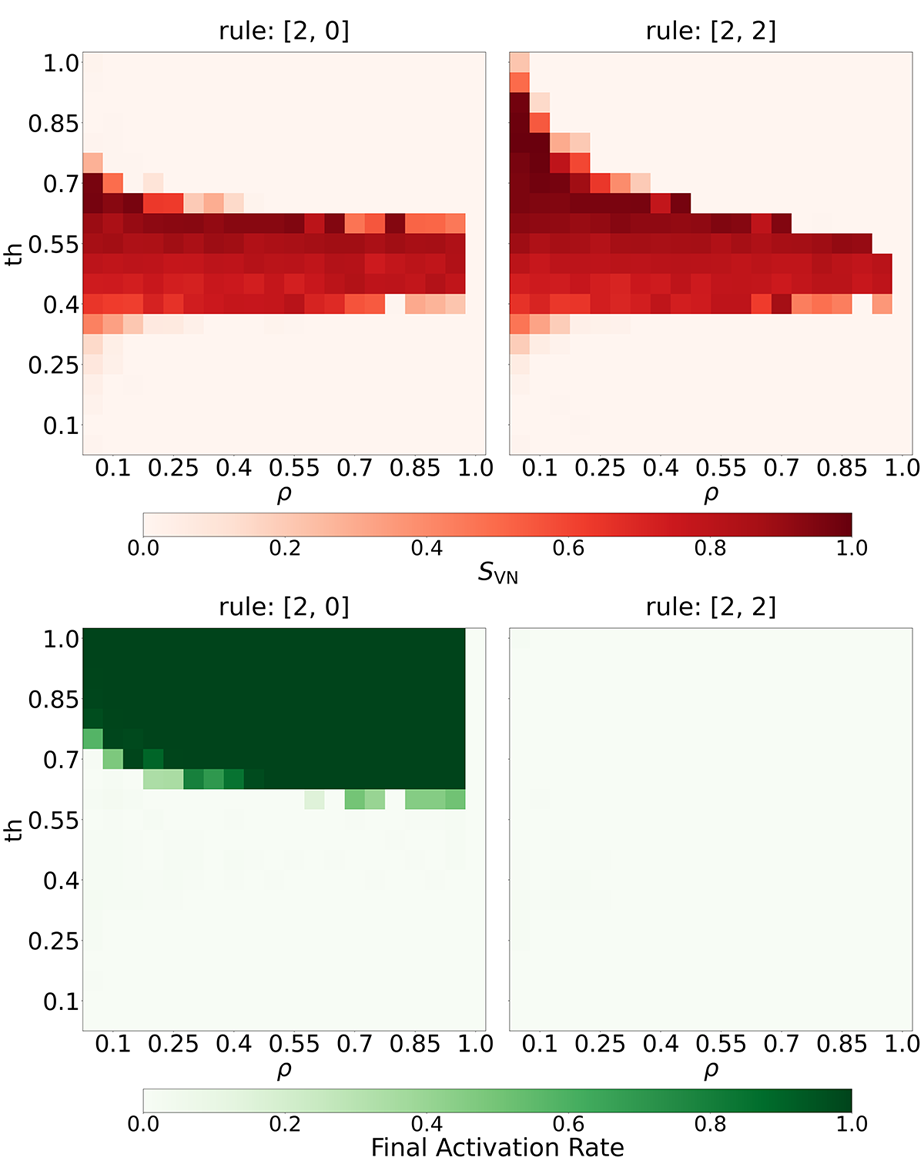}
\caption{\label{fig:PDrationality} Coordination failures of the Class 2 Phase Diagram. Results are shown for agents employing random search ($s=2$). Seeking unconditional availability (left: $m=0$) yields dynamic coordination failure at high thresholds (top row phase diagrams), characterized by a high final activity rate for $\text{th}>\vartheta_m$ (bottom row phase diagrams). Seeking satisfaction (right: $m=2$) yields static coordination failure as no sites in a random initial distribution meet the strong homophily demand. All simulations run with Rivalry on, $L=100$ and $r=10$.} 
\end{figure}

\begin{figure}
\includegraphics[width=8.7cm]{ 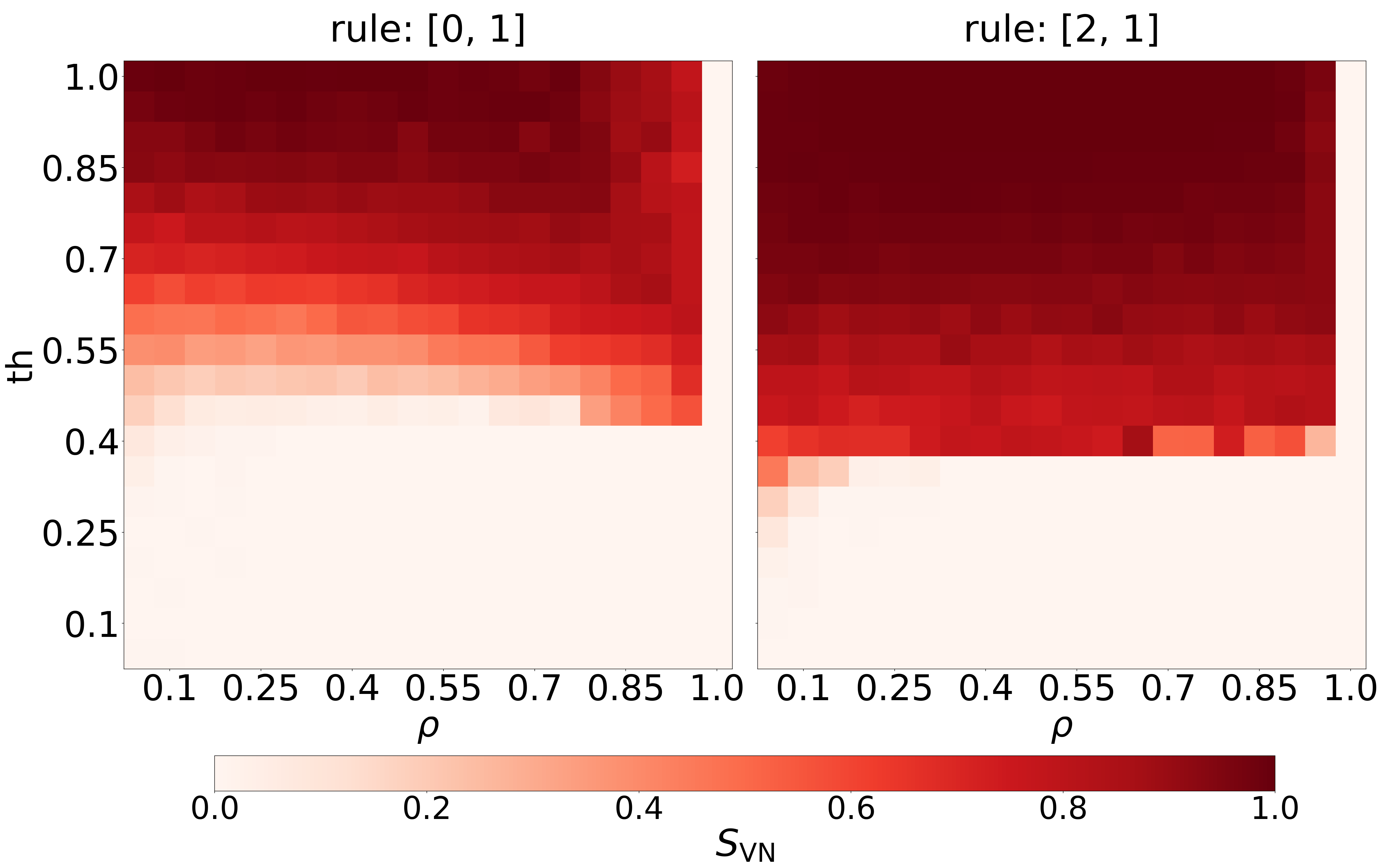}
\caption{\label{fig:move-improved} Class 1 phase diagrams. An `improvement seeking` movement criteria ($m=1$) enables agents to reinforce and amplify fluctuations in the initial state to attain strong homophily goals, resolving Class-2 coordination problems to yield Class-1 phase diagrams. Results shown for NN search (left: $s=0$) and random search procedures (right: $s=2$). All simulations run with Rivalry on, $L=100$ and $r=10$.} 
\end{figure}

This relaxation of the originally pathological utility function resolves the coordination issues seen in Class-2 (Figure \ref{fig:PDrationality}), leaving only the primary Schelling transition at $\vartheta_s$ that characterizes Class-1 (Figure \ref{fig:move-improved}). Notably, this  refinement parallels similar trends in the decision theory literature contemporaneous with Schelling, such as Simon's 1956 work on ``satisficing'' behaviors \cite{simon1956rational} that allow for good-enough optimization in essentially noisy biological systems \cite{bialek2012biophysics}. As such, continuity is now considered a basic aspect of well-behaved utility functions \cite{mas1995microeconomic}.  

\subsubsection{\label{subsubsec:restless}Restless Rascals}
When agents ignore satisfaction goals and
move as long as the site is available (i.e., vacant) at high thresholds, most agents are unsatisfied and thus relocate. However since these agents ignore satisfaction goals, the system shuffles through random initial states each with $q \sim 0.5 \pm \sigma_q$ and agents are thus equally unlikely to be satisfied wherever they relocate. The few agents who do stumble into satisfactory locations are quickly perturbed by the random relocation of the remaining agents. This results in continuous stochastic motion due to a failure to dynamically coordinate and settle, evidenced by the dynamic activity of the final state.

To measure the dynamics at convergence, the average fraction of agents that relocated every 20 steps over the last $2N$ steps (see Appendix \ref{sec:Conv}) was used as an indicator of activity levels, this phase diagram is shown for the Class-2 subclasses in the second row of Figure \ref{fig:PDrationality}. 

\subsubsection{\label{subsubsec:petrified}Petrified Purists}

Under the step utility function (\ref{eq:utility}), agents who relocate with preference for satisfaction (m=2) are restricted to relocating to available sites with $q \geq \text{th}$. At high thresholds, neighborhoods that deviate meaningfully from the ensemble average of $q=0.5$ become exceedingly rare when neighborhoods are of moderate size (see Figure \ref{fig:qsame histogram}). The few agents that do move are apparently insufficient to initiate large-scale cascades, resulting in minimal overall change. Consequently, the final configuration remains largely, if not entirely, similar to the initial one. This static coordination failure traps agents in a frustrated state, where their preferences cannot be satisfied by individual action. This also explains the $\rho$ dependence of $\vartheta_m$. As the lattice gets fuller, fewer satisfactory sites (havens) are available even at moderately high thresholds, thus lowering $\vartheta_m$ as $\rho$ increases. This frustrated phenomenon is verified by calculating the threshold at which the availability of havens becomes insufficient (see Figure \ref{fig:havens}) and examining the lack of dynamic activity at convergence (see Figure \ref{fig:PDrationality}).
\subsection{\label{subsec:class0} Understanding Class 0}
\begin{figure*}
\includegraphics[width=13cm]{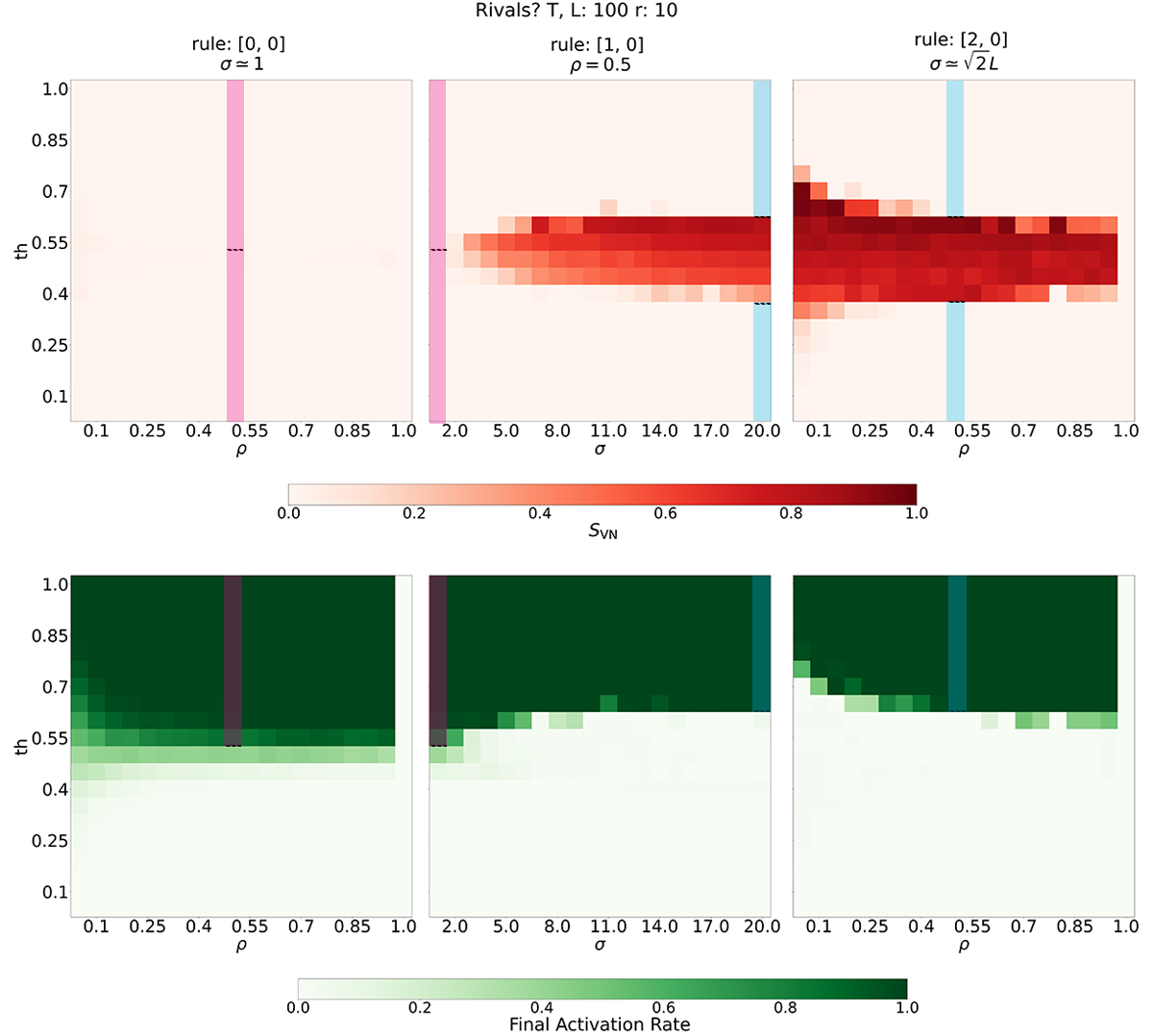}
\caption{\label{fig:class0 exp} Comparison of Class 0 and Class 2 Phase Diagrams. A distance-weighted search procedure (center column) interpolates between nearest neighbor (left column) and random (right column) search procedures. The top row shows the segregation phase diagrams for nearest neighbor [0,0] and random [2,0] versus density with the distance weighted search [1,0] versus search radius between them. The corresponding columns between search procedures are marked by colored bars. All three are characterized by high dynamic activity at convergence (bottom row). Despite the nearest neighbor search outcomes appearing as a unique segregation class (Class 0), it exhibits the same macroscopic behavior as random search (class 2), differing only in the location of the upper transition as a result of prospecting vision. Distance-weighted search shows this transition lifting smoothly as prospecting vision increases. All simulations run with Rivalry on, $L=100$, and $r=10$.}
\end{figure*}
As shown in Figure \ref{fig:class0 exp}, class 0 can be seen as an edge case of the Class 2 phase diagram. In particular, $\vartheta_s, \vartheta_m$ both converge to 0.5 due to constrained prospecting vision and a blind movement rule. The segregated region is ``pinched out'' of the phase diagram as these transitions overlap. Constrained prospecting vision paired with a large neighborhood increases $\vartheta_s$ due to localization of the avalanche and increasing correlations between the dissatisfied agent's homophily quotient and that of the prospect sites. Thus, any improvements achieved (despite the blind movement criteria) are minimized and the transition threshold is raised to $\vartheta_s \simeq 0.5$. The upper transition at $\vartheta_m$ is also lowered to $\vartheta_m \simeq 0.5$ as the `blind' movement criteria favors picking the closest available site which will have a high degree of correlation to the originally dissatisfying $q$ and is thus unlikely to offer satisfaction even at moderate th.  Ultimately, at low thresholds with constrained prospecting vision, not enough seeds are available to trigger the avalanche. Once enough seeds become available the `blind' move rule results in dynamic coordination failures at lower thresholds than usual as a result of the constrained prospecting vision.

While no transition can be seen in segregation outcomes, a transition in the dynamics of agents can be seen at this point of overlap. Furthermore, the distance weighted search still interpolates smoothly between the NN and random rule sets under move always (see Figure \ref{fig:class0 exp}), confirming it has equivalent behavior to that seen under (s,m) = (2,0) but with the transitions at ($\vartheta_s, \vartheta_m$) overlapping.


\subsection{\label{subsec:vacancies}Role of Vacancies}

In Schelling's original model, agents compete for vacancies. In such a model, vacancies play two pivotal roles. Most obviously, vacancies facilitate mobility as agents can only relocate into vacant sites. More subtly, vacancies regulate the expected number of agents in a neighborhood (see equation \ref{eq:nbhood density}). Neighborhoods with a higher proportion of vacancies contain fewer agents, and are more sensitive to being destabilized by immigration and emigration. Consequently, the effects of relocation on local neighborhood composition are amplified---as empirically verified in Figure \ref{fig:fluctuations-all}. However, in this case, the reduction in  $\langle n \rangle $ and thus increase in $\sigma_q$ is driven by an decrease in occupation density $(\rho)$ rather than a smaller neighborhood radius.

To isolate the two functions of vacancies, a rivalry parameter is introduced. In all previous phase diagrams rivalry was enabled, with agents competing for vacancies as sites are restricted to single agent occupancy. When rivalry is turned off, multiple agents can occupy the same site, effectively turning off the role of vacancies in facilitating movement. The resulting phase diagrams are explored below.
\subsubsection{\label{subsubsec:blocking}Multi-agent occupation}
Enabling multi-agent occupancy affects the upper transition for agents with preference for satisfaction (m=2). When a haven is available, these agents preferentially pile into this site, often generating more havens in the process. As a result, segregation is now possible at higher occupation densities and thresholds than in the single-agent occupancy case. 

However, this effect has a limit: for the avalanche to begin, agents must be mobile and thus at least one haven must exist in the initial state. At sufficiently high thresholds and occupation densities, no haven is available and agents remain trapped in their initial positions. The cutoff threshold at which no havens are available is calculated as follows.

\begin{figure}
\includegraphics[width=8.7cm]{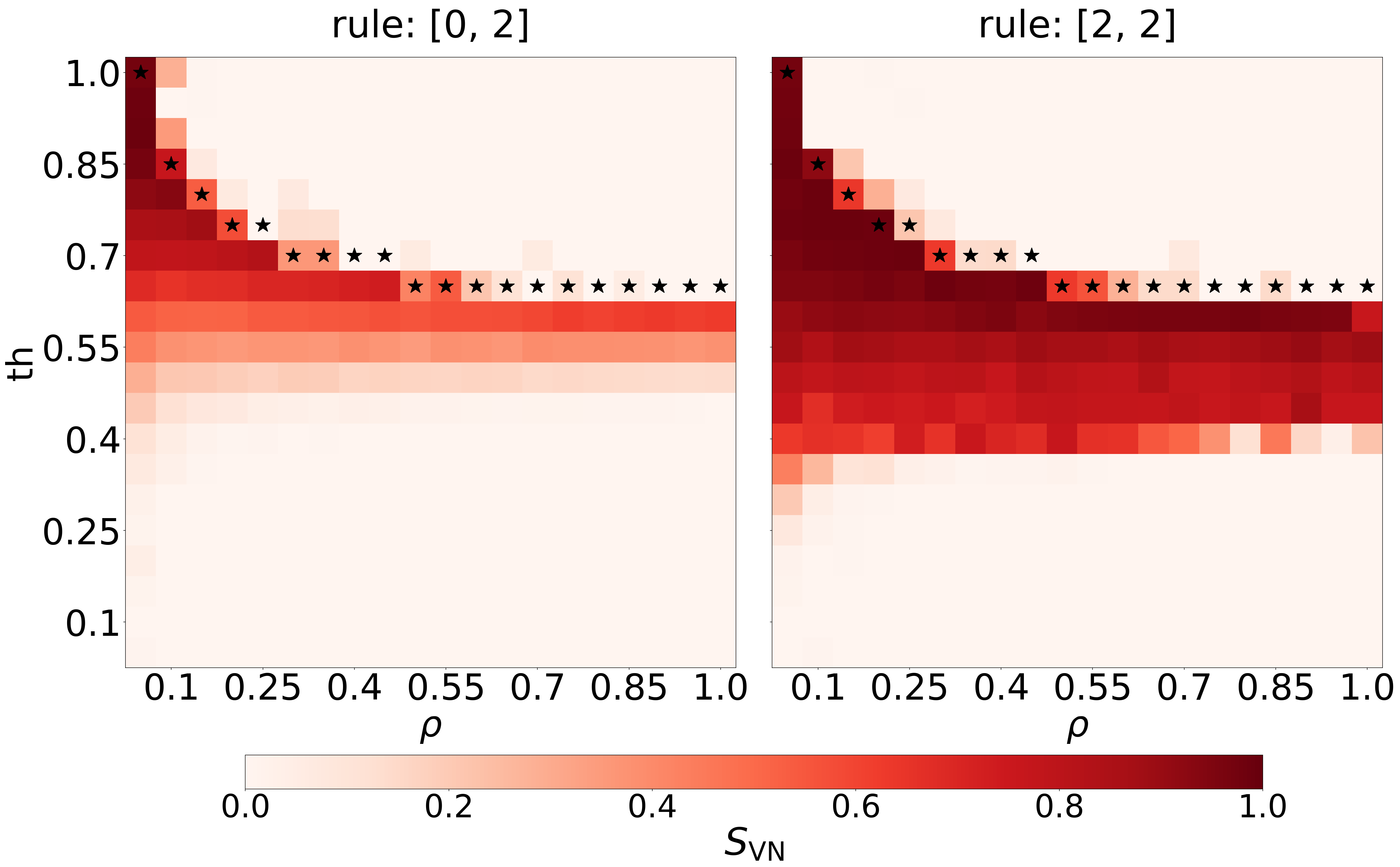}
\caption{\label{fig:havens} Class 2 outcomes for satisfaction driven agents when multi-agent site occupation is permitted. Stars indicate the cutoff threshold where the expected number of havens first drops below 1. The upper transition for satisfaction seeking Class 2 phase diagrams matches the drop-off in availability of havens, for both nearest neighbor (left) and random (right) search procedures. All simulations were run with Rivalry off, $L=100$ and $r=10$.}
\end{figure}

First, the probability that a given neighborhood has $n$ agents is multiplied by the probability that $n_s$ of those agents are of the same type and $n_o = n-n_s$ are of the opposite type. Both probabilities can be calculated with the binomial probability distribution.
\begin{eqnarray}
p(n, n_s) = p(n)p(n_s, n-n_s)
\label{eq:likelihood of nbhood}
\end{eqnarray}
The probabilities of events associated with agent satisfaction are then summed to determine the total probability of an agent finding a neighborhood to be satisfactory. We refer to a satisfactory neighborhood as a haven. 
\begin{align}
p_{\rm haven}(N, \text{th}) = \sum_{n=1}^{N}p(n) &\sum_{\frac{n_s}{n} \geq \text{th}} p(n_s, n-n_s) \nonumber \\
& + p(n=0)
\label{eq:likelihood of haven}
\end{align}
For a given occupation density, the cutoff threshold is the threshold at which the expected number of havens first drops below 1:
\begin{eqnarray}
\mathbb{E}_{\rm havens}(N, \text{th}_{\rm cutoff}) < 1.
\label{eq:exp havens}
\end{eqnarray}
This was calculated at each occupation density and is indicated by a star in Figure \ref{fig:havens}. This cutoff closely matches the observed upper transition, confirming its origin in frustrated initial states.

For agents who relocate without a preference for improvement/satisfaction ($m=0$), rivalry had no effect. Since these agents relocate blindly, they are unable to coordinate and preferentially pile into satisfactory sites, preventing the formation of additional havens. Ultimately, stochastic motion persists as few, if any, agents settle into satisfactory locations, and those that do are often quickly perturbed by the relocation of others. 
\subsection{\label{subsec:seg measures}Measures of Segregation}

Schelling originally measured segregation in his 1-D model using neighborhood statistics such as the distribution of cluster sizes and the proportion of agents with no neighbors of the opposite color. In his 2-D model he introduced the average proportion of like neighbors ($Q$) as a measure of neighborhood homogeneity. The Von-Neumann segregation metric $S_{\rm{VN  }}$ is a normalized version of the ensemble averaged homophily quotient $Q$. Methodological analysis by Fossett \cite{Fossett} further supports its use, showing that  $S_{\rm{VN}}$ is equivalent to Bells revised index of isolation \cite{Bell} and strongly correlated to the eta-squared index($\eta^2$) and the index of dissimilarity (D), each of which is a commonly used metric for segregation in sociology.

Since changes in the order parameter are central to the classification scheme, it is important to assess the suitability of the segregation metrics employed—and to investigate whether alternative measures may reveal previously unobserved classes of behavior. Alternative segregation measures inspired by percolation theory were explored and are defined in section \ref{subsubsec:perc}. Importantly, the classification scheme was found to be robust to the choice of segregation measure, with the distribution of classes and character of each class remaining unchanged.

\subsubsection{\label{subsubsec:perc}Percolation-Inspired Segregation Measures}
Schelling emphasized that clustering measures are related to---but distinct from---measures of neighborhood homogeneity. Consequently, some analyses have incorporated segregation measures inspired by percolation theory. These study the geometry and statistics of clusters, where two agents belong to the same cluster if they are nearest neighbors. Clusters are labeled and identified using the Hoshen-Kopelman algorithm\cite{HK-Alg}. Due to clusters being defined by the four NNs, the presence of vacancies can yield poor identification of visually identifiable clusters. This is amended with a bootstrap process (introduced in \cite{PDSchel}, where they refer to it as ``renormalization''), modified such that each vacancy is filled by performing a majority vote of its nearest  neighbors. For example, a site surrounded by 2 blue agents, a vacancy and a red agent shall be filled with a blue agent. Similarly a site surrounded by 3 vacancies and a red agent is left vacant. For all of the cluster-based metrics considered, the lattice is bootstrapped prior to cluster identification.

Cluster geometry is typically characterized by averaging geometric properties over all identified clusters $\{c\}$ on the lattice. Common measures of geometry include cluster mass ($m_c$) and radius ($r_c$). Alternatively, non-averaged quantities that decrease as segregation emerges can be used to capture the degree of agglomeration in the system. The total number of clusters ($\nu$) is a natural candidate: in maximally segregated states, clusters grow larger, thus reducing the total number of clusters. Segregated states also minimize contact between opposite agents types. Thus the total number of opposite-type nearest-neighbor pairs will be minimized relative to the total number of NN pairs. This quantity, known as the interfacial density \cite{UnifiedSchel} captures agent exposure to the opposite type by measuring the fraction of frustrated bonds. 

All of these measures were implemented and found to reproduce the distribution of the 3 classes identified with no changes in the location or number of transitions (see Figure \ref{fig:seg measures} in Appendix \ref{sec:Measures}). Details of how these measures are calculated are given in Appendix \ref{sec:Measures}


\section{Future Outlook\label{sec:Outlook}}
We have systematically studied a large variety of rule sets implementing the conceptual Schelling model. Many of these different instantiations of the same broad idea can be found in the literature, but little had been done to check that these all represent the same model in terms of the important behaviors. We find that the 54 rule variants we considered collapse down to three classes, characterized by qualitatively different segregation/integration phase diagrams of the models. Two of these classes exhibit seemingly pathological behaviors, while the third captures the surprisingly low threshold for segregation highlighted in Schelling's original work. Since many important works in the Schelling literature use rule variants from at least two of these classes, careful consideration is needed when synthesizing results from across the literature.

We also presented two important insights into the nature of the integration to segregation phase transition. We find that the key to understanding the Schelling transition lies in the initial statistics of neighborhoods. Segregation in the Schelling class of rules comes from a small number of dissatisfied agents who move early on and in the process seed dissatisfaction in their new neighborhoods. This starts a chain reaction that causes the model to segregate. We find clear effects from density and neighborhood size that point to this avalanche seeding process. Additionally, by considering models with and without rivalry, we're able to separate two distinct roles of vacancies in the Schelling model: They increase mobility in models with rivalry and they change the initial statistics of neighborhoods. Previous studies have focused on the former, but we find that added mobility only has a strong impact in some cases with unrealistic parameter values or rule sets from the pathological classes. 

This points to a variety of directions for further study. We see much more realistic behavior when we use a continuous utility function rather than the step function implied by Schelling's original rules (i.e., ``move improved'' versus ``move satisfied''). This suggests a clear short-term question: How do differently shaped utility functions affect the phase diagram? This question was first addressed by Pancs et al \cite{pancs2007utility} and is currently being explored in light of our present results.

More broadly, connecting Schelling's original insight to real-world data, in the housing sector or in a variety of other contexts that have been suggested in the literature \cite{randall2022het, bruch2014welfare, axtellAS}, requires complicating the model. These complications fall broadly into the categories of complicating the agents (adding extra properties that affect movement), making the agent populations heterogeneous, and making the environment heterogeneous (with some connected complications of the agents and their interactions with such an environment). We refer to this broad class of models that can grow out of the simple Schelling model as positional choice models. Our work suggests that systematic variation of parameters and rule variants will be important to understanding mechanisms in more complicated positional choice models, since it is already providing valuable insights into the simple, well-studied Schelling model.

\begin{acknowledgments}
We thank Lyndsy Acheson for her assistance in testing the code for violations of the decision rules and search procedures. We also thank Steve Weiss and the Institute for Quantum Computing (IQC) for providing remote servers for running the 54 rule variants.  This research was funded by the University of Waterloo, IQC, and NSERC. 
\end{acknowledgments}
\appendix
\section{\label{sec:Measures}Measures of Segregation}

\begin{figure*}
\includegraphics[width=17.5cm]{ 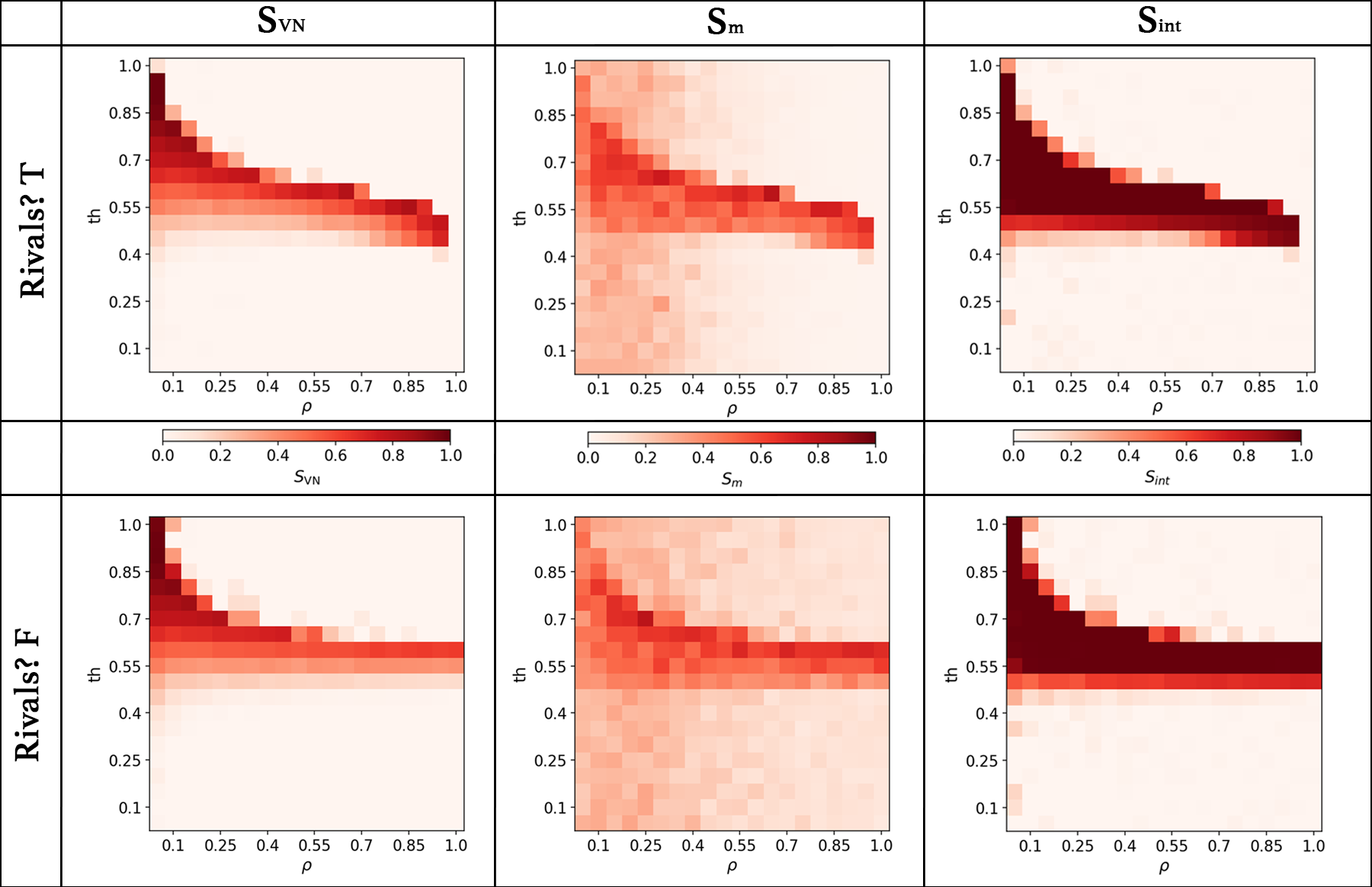}
\caption{\label{fig:seg measures} Comparison of segregation measures. Results shown for satisfaction seeking ($m=2$) Class 2 phase diagrams under NN search ($s=0$). Von-Neumann segregation measure (left column), Weighted average mass (middle column) and interfacial density segregation measure (right column). The small degree of clustering that occurs at moderate occupation densities acts as background noise, obscuring the distinct phases present in the $S_m$ phase diagram (middle column). All simulations run with $L=100$ and $r=10$. Top row is with Rivalry on, bottom row is with Rivalry off. } 
\end{figure*}
We considered a variety of segregation metrics coming from different fields and found that they all give the same phase diagrams as the von Neumann segregation used in the paper (see Figure \ref{fig:seg measures}). Gauvin et al.\cite{PDSchel} introduce a weighted average of the \textbf{cluster mass}:
\begin{eqnarray}
S_{m} = \frac{2}{N} \sum_{\{c\}} m_c w_c = \frac{2}{N^2} \sum_{\{c\}} m_c^2,
\label{eq:avg cluster mass}
\end{eqnarray}
where each cluster $c$ contains $m_c$ agents, and is weighted by the probability that an agent belongs to cluster c: $w_c = m_c/N$. This is then normalized by the maximum cluster mass $N/2$. A system dominated by large clusters will thus yield $S_m$ close to 1, and a system dominated by small clusters will yield $S_m$ close to 0. 

It is important to note that while the bootstrapped lattice is used for labeling clusters, these `ghost' agents that were filled in are not counted towards the cluster mass. 

This measure reproduces the original classification results, however it also registers the small degree of clustering present in a random distribution of agents at low occupation densities where $\sigma_q$ is large. This small degree of clustering appears as background noise present in the integrated regions of the phase diagram. When rivalry is off and the density of occupied sites remains around 0.5, the background noise registers as flat across the integrated regions of the phase diagram.   

\begin{figure*}
\includegraphics[width=13cm]{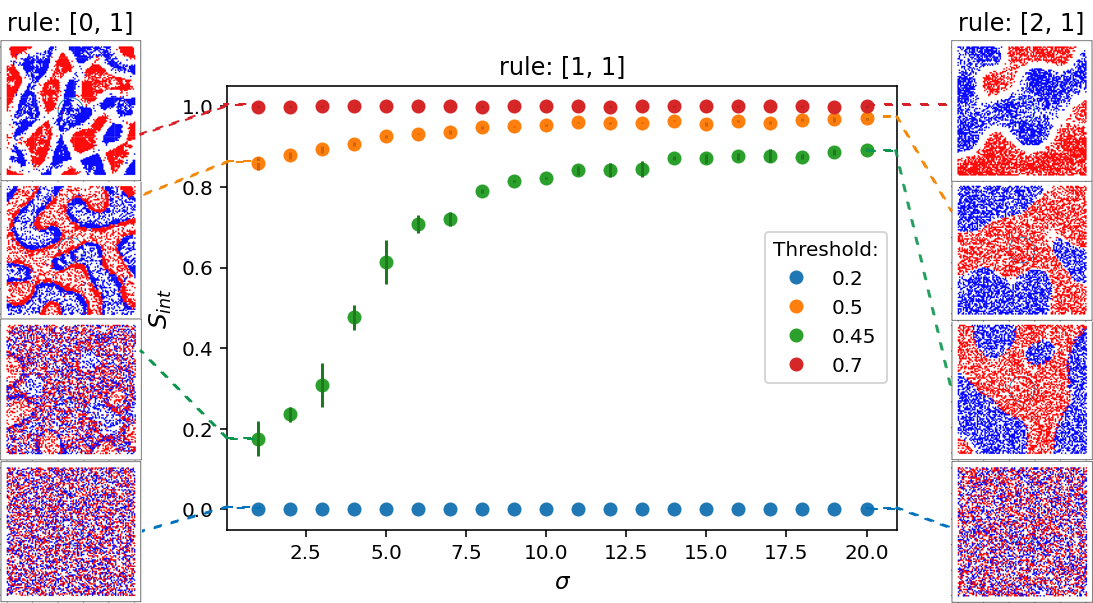}
\caption{\label{fig:DW-connection} Distance Weighted Search interpolates between the limiting cases of local and global vision. For $\sigma = 1$ distance weighted search results in an interface density matching that obtained under NN search. For $\sigma = 20$ the interface density matches that obtained for random search. Similar behavior is seen for other movement criteria $m=0, 2$.}
\end{figure*}

In a maximally segregated outcome, agents minimize exposure to agents of the opposite type (\textbf{interfacial contact}). Analysis of edges (in the graph theoretic sense, i.e., NN pairs) can thus be used to calculate interfacial contact and generate a segregation measure. The interface density $\delta$ is calculated as the number of opposite edges ($\epsilon_{AB}$), divided by the total number of edges containing at least one A ($\epsilon_{AA} + \epsilon_{AB}$):
\begin{eqnarray}
\delta = \frac{\epsilon_{AB}}{\epsilon_{AA} + \epsilon_{AB}},
\label{eq:int_density}
\end{eqnarray}
Since $\delta$ will be minimized for segregated states and maximized for integrated states, the proximity to complete integration is measured. For a completely integrated lattice, it is equiprobable for an edge to be between opposite type agents as it is for an edge to be between same type agents and thus $\delta_{\rm int} = 0.5$. Thus a normalized measure can be determined as:
\begin{eqnarray}
S_{\rm int} = 1-  \frac{\delta}{\delta_{\rm int}}
\label{eq:S_int}
\end{eqnarray}
Where as $\delta \to \delta_{\rm int}$, $S_{\rm int} \to 0$ and as $\delta \ll \delta_{\rm int}$ due to segregation, $S_{\rm int} \to 1$. In Figure \ref{fig:DW-connection}, the interface density-based segregation measure $S_{int}$ was used to measure the segregated outcomes of distance weighted search, connecting the clustering behavior observed for nearest neighbor search to that seen for random search.

The \textbf{radius of gyration} $r_c$ \cite{Perc} for a cluster $c$ of mass $m_c$ is derived from the variance in the separation from the center of mass $r_o$:
\begin{eqnarray}
r_c^2 = \frac{1}{m_c} \sum_{i=1}^{m_c} |r_i-r_o|^2.
\label{eq:gyr_radius}
\end{eqnarray}
Where the center of mass $r_o$ is:
\begin{eqnarray}
r_o = \frac{1}{m_c} \sum_{i=1}^{m_c} r_i
\label{eq:COM}
\end{eqnarray}
The ensemble averaged radius of gyration is then compared to the radius of gyration achieved by complete segregation $r_s$
\begin{eqnarray}
S_r = \frac{1}{r_s} \sum_{\{c\}} r_c.
\label{eq:radius_measure}
\end{eqnarray}
Where $S_r \to 0$ as the clusters get smaller (integrated outcomes) and $S_r \to 1$ as the clusters get larger and approach the fully segregated limit.

As the lattice maximally segregates, clusters get larger and larger, minimizing the \textbf{number of clusters} ($\nu$). Thus the number of clusters present in the final state $\nu$ can be compared to the number expected in an integrated state $\nu_{\rm int}$. Where $\nu_{\rm int}$ is calculated numerically by averaging 100 random initial states. A measure is thus calculated as:
\begin{eqnarray}
S_{\rm count} = 1-\frac{\nu}{\nu_{\rm int}},
\label{eq:cluster count}
\end{eqnarray}
Where $S_{\rm count} \to 1$ as segregation emerges. 

That each of these various segregation metrics agreed in the phase diagram validates the utility of phase diagrams for analyzing the structure of Schelling model outcomes and dynamics.

\section{\label{sec:Conv}Convergence Testing}

A convergence metric is required to determine when the simulation has reached a steady state and should terminate. The convergence criterion should capture either the absence of updates (static convergence), or the presence of updates that no longer alter the macroscopic state of the system (steady state convergence). To evaluate steady state convergence, a relevant system metric is recorded over time. Every $2N$ steps, the stored dataset of these metrics is split into two halves, and the difference in their means is compared against the expected statistical fluctuation:
\begin{figure}
\includegraphics[width=8.5cm]{ 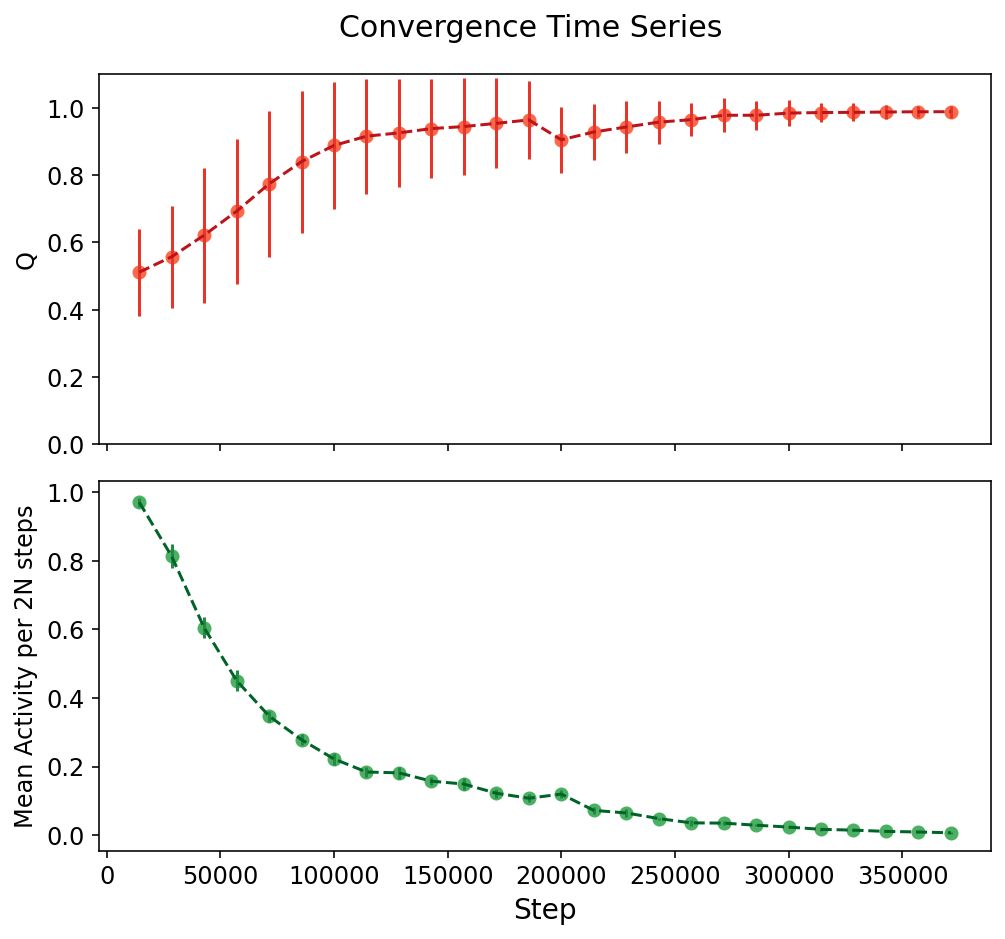}
\caption{\label{fig:tseries} Sample of a simulation timeseries. Convergence timeseries shown for a simulation with th=0.9, $\rho=0.7$, $r_f=0.1$, $L=100$ executing improvement seeking relocations under the NN search procedure. Simulation terminates when condition \ref{eq:conv-condition} is satisfied for $Q$ or the activity rate drops to 0.} 
\end{figure}
\begin{eqnarray}
|\mu_1 - \mu_2| < C \sqrt{\frac{\sigma_1^2 + \sigma_2^2}{N}},
\label{eq:conv-condition}
\end{eqnarray}
Here, $\mu_1$ and $\mu_2$ are the means of the first and second halves of the dataset and $\sigma_1$ and $\sigma_2$ are the variances. $C$ is a tunable confidence parameter chosen to be $1/4$ for all phase diagrams run. 

Ideally, due to the large number of steps needed for a simulation to converge, calculating this metric and recovering its results should be computationally inexpensive. Since each agent's homophily quotient is already calculated at every step, it serves as the primary steady state convergence metric. The simulation is terminated once this measure of neighborhood homogeneity has stabilized. 

Alternatively to detect static convergence, a ticker is incremented every time an agent relocates during a step. Every 20 steps, the ticker is used to calculate the fraction of agents probed that actually relocated, before being reset to 0. The average of this `activity log` is then checked every $2N$ steps alongside homophily convergence. If the average activity in the last $2N$ steps is 0 (as is the case in Figure \ref{fig:tseries}), then the simulation is flagged as statically converged and terminates. The average activity in the final $2N$ steps may then be used as an order parameter as seen in the second rows of Figures \ref{fig:PDrationality} and \ref{fig:class0 exp}.

\section{Minimizing Uncertainty}

It is important to minimize the uncertainty in each data point of the segregation phase diagram. As noted in sections \ref{subsubsec:radius} and \ref{subsec:vacancies}, the final outcome is sensitive to the initial distribution. Close to the phase transitions the outcome is especially sensitive to the presence of avalanche dissatisfaction seeds in the initial state. 

To average out any uncertainty due to the initial state, each data point is run at least 3 times, each with different initial states in which agents are distributed randomly. The standard deviation in the runs' final average homophily $Q$, is used to determine if a given data point requires additional runs. (See \cite{PDSchel} for examples of the distribution of outcomes). Data-points that had a distribution of outcomes with a standard deviation greater than 0.03 are identified and ran 17 more times (for a total of 20 runs) in order to calculate a trimmed (20\%) standard deviation. This was done to differentiate data points whose distributions contained outliers, from those with bimodal outcomes. It was found that data points which still had a standard deviation greater than 0.15 had bimodal outcomes with final states either being segregated or remaining close to the initial state.

\section{Role of Prospecting Time\label{sec:Prospecting}}
\begin{figure}
\includegraphics[width=8.7cm]{ 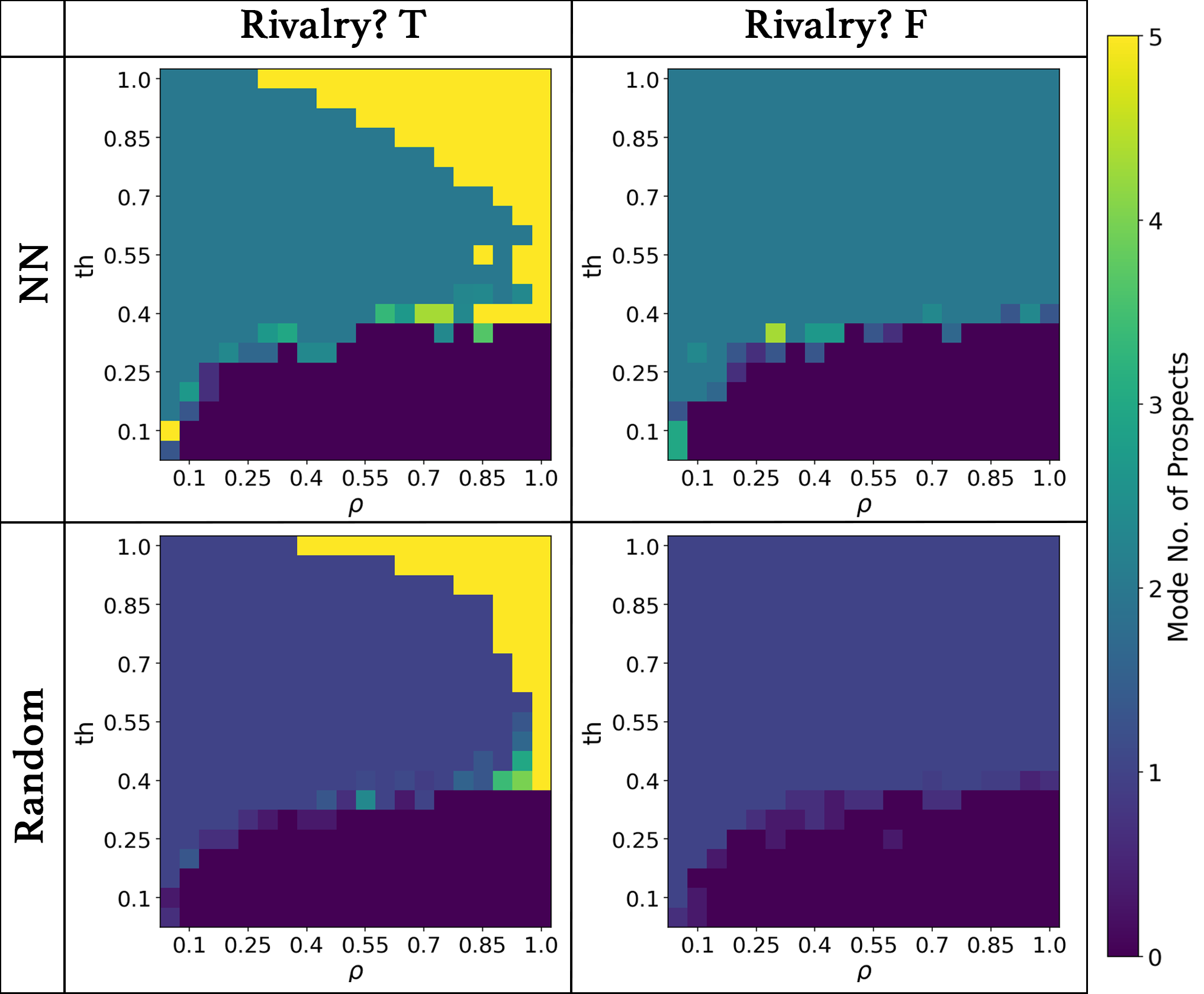}
\caption{\label{fig:prosp-improved} Number of sites prospected for improvement seeking agents. Under nearest neighbor search (top row: $s=0$) agents searching for improvement ($m=1$) must prospect more sites in comparison with random search (bottom row: $s=2$). Additionally, when site rivalry is on (left column) prospecting time approaches $\ell_p$, which is set to its upper-bound, due to limited availability of vacancies offering improvement. The colorbar maximum was set to 5 to enable resultion in the middle region of the phase diagrams. All simulations run with $\ell_p = (L^2-1)$, $L=100$ and $r=10$.} 
\end{figure}
The loop constraint $\ell_p$ sets an upper limit on the the number of sites an agent can prospect before giving up. If a site satisfying movement criteria is found in under $\ell_p$ attempts, the agent relocates; otherwise, they remain at their current, unsatisfactory location. In Schelling's original checkerboard model, agents prospect until a satisfactory site is found, effectively imposing no upper-limit on the number of sites prospected ($\ell_p = \infty$). Under move-improved (m=1) or move satisfied (m=2), this would result in long run times when few suitable sites are available. To avoid this, if $m>0$ and $\ell_p = \infty$, the list of prospect sites is pre-filtered to include only those meeting the agent’s movement criteria, under the assumption that the agent would eventually find any qualifying site on the lattice. All phase diagrams in the main body of the paper are run with $\ell_p = \infty$. 

\begin{figure}
\includegraphics[width=8.7cm]{ 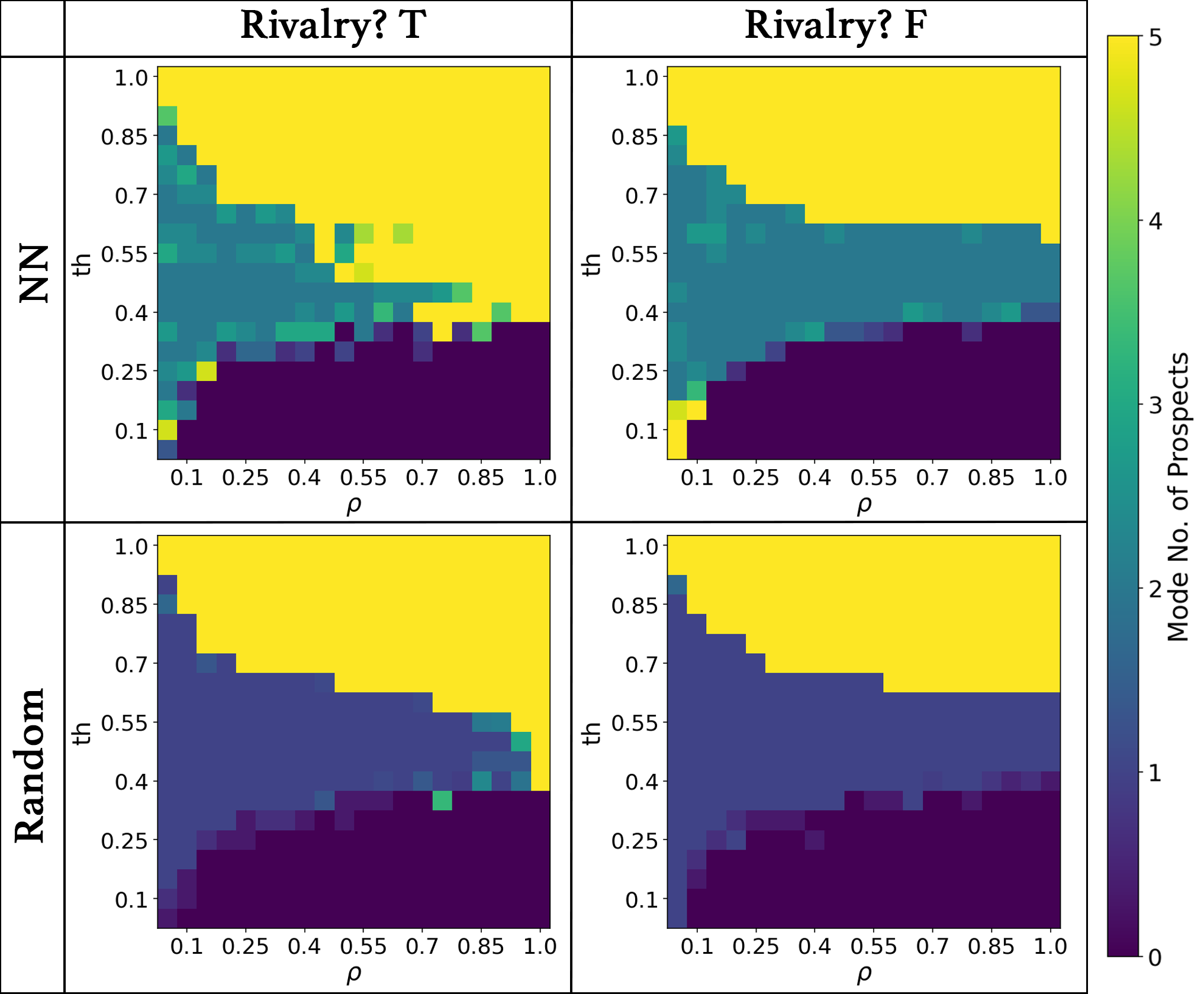}
\caption{\label{fig:prosp-satisfied} Number of sites prospected for satisfaction seeking agents ($m=2$). NN search (top row: $s=0$) agents searching for satisfaction must prospect more sites in comparison with random search (bottom row: $s=2$) for th $>\vartheta_s$. For both rivalry conditions, prospecting time approaches $\ell_p$, which is set to  its upper bound, for th $> \vartheta_m$ as few to no satisfactory sites are present in the lattice. Colorbar maximum was set to 5 for resolution in th $< \vartheta_m$ region. All simulations run with $\ell_p = (L^2-1)$, $L=100$ and $r=10$.} 
\end{figure}

To understand the effect of prospecting time, the lower limit of $\ell_p=1$ is also tested. Additionally, a finite $1 < \ell_p < \infty$ is set in order to study the mode number of prospects agents need when searching without pre-filtering of prospects. 
\subsection{Finite Limit: $1< \ell_p < \infty$ \label{subsec:Prospecting-finitelimit}}

The finite limit on $\ell_p$ was set to $L^2-1$. Under this constraint, no prefiltering of prospects takes place and agents must inspect each site for satisfaction of movement criteria. For improvement seeking agents, a large upper limit on prospecting did not change segregation outcomes despite agents prospecting closer to this limit when th and $\rho$ approach 1. Examination of the mode number of sites prospected in each region of the phase diagram (see Figures \ref{fig:prosp-improved}, \ref{fig:prosp-satisfied}) further confirms the role of vacancies in enabling mobility and highlights the prospecting efficiency of random search in comparison to NN search.
\subsubsection{Move Improved\label{subsubsec:finitelimit(m=1)}}

For nearly all ($\rho$, th) agents are able to find sites that offer an improvement within only a few prospects. The specific number of prospects required depends on the efficiency of the search algorithm. Random search was more efficient than NN search, with most agents only prospecting 1 site before successful relocation. NN search required an extra prospect, with most agents prospecting 2 sites before relocating. 
Despite agents only seeking improvement, when rivalry is enabled and occupation density is high, most sites are unavailable. When combined with intense homophily demands, this causes the majority of prospected sites to be non-viable, significantly increasing the number of sites an agent must prospect. When rivalry is turned off, this increase in number of sites prospected is no longer seen (see Figure \ref{fig:prosp-improved}). 

\subsubsection{Move Satisfied\label{subsubsec:finitelimit(m=2)}}
\begin{figure}
\includegraphics[width=8.7cm]{ 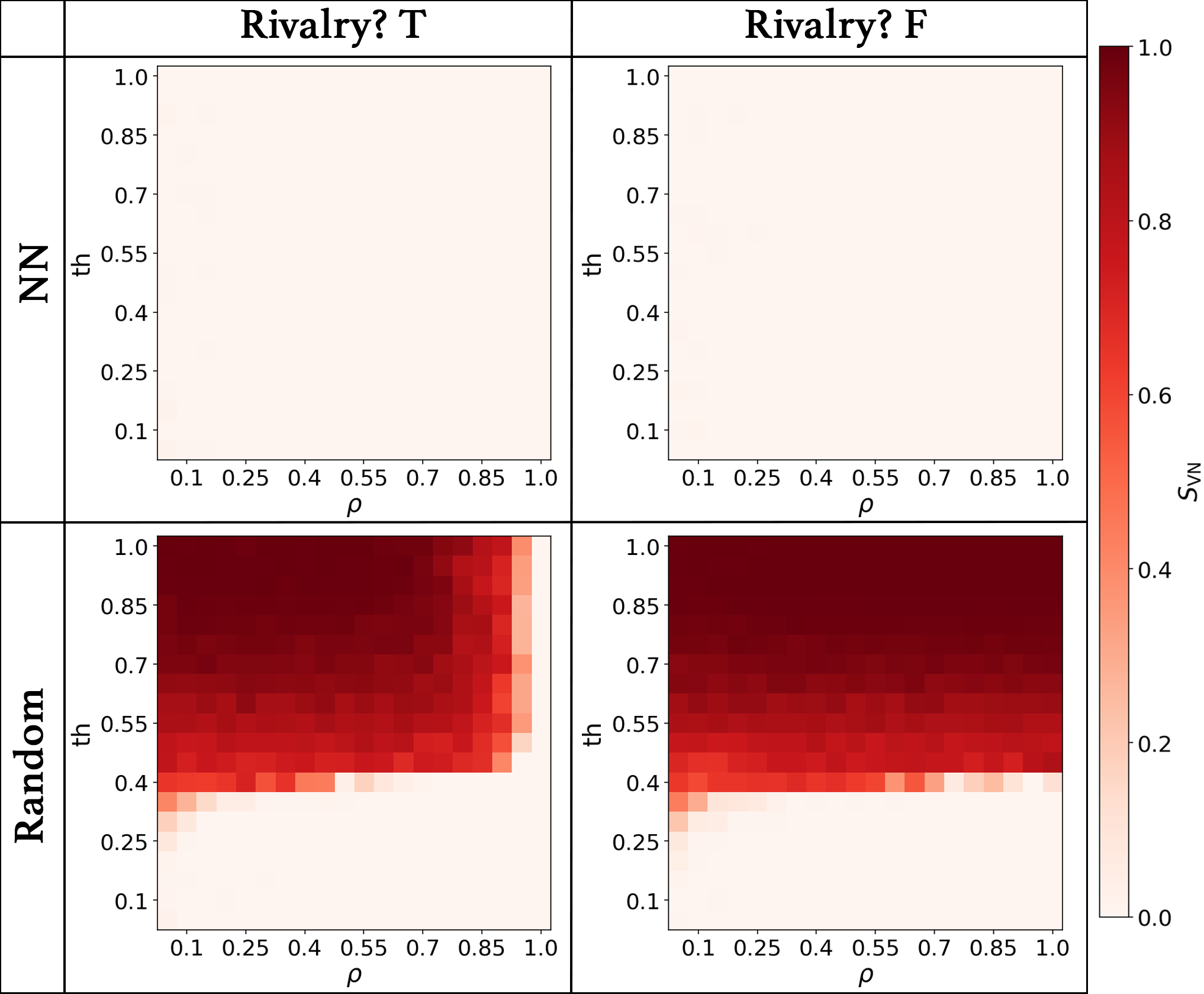}
\caption{\label{fig:prosp1-improved} Segregation outcomes of constrained prospecting. Segregation outcomes for improvement seeking agents when prospecting is constrained to 1 site. NN phase diagrams collapse to Class 0 as agents require $>1$ prospect due to inefficient search procedure. All simulations run with $\ell_p = 1$, $L=100$ and $r=10$.} 
\end{figure}
At moderate homophily demands ($\text{th} < \vartheta_m$), the number of sites prospected for satisfaction is comparable to the number of sites prospected for improvement. This is because at moderate th many havens are available. For $\text{th} > \vartheta_m$, havens become scarce, leading to a sharp increase in the number of sites prospected. Once no havens are available, agents consistently reach the prospecting limit before giving up. 

\subsection{Lower Limit: $\ell_p = 1$ \label{subsec:Prospecting-lowerlimit}}
When agents are limited to prospecting a single site, inefficient search algorithms that require multiple prospects fail to identify viable relocation options. This results in integration at moderate thresholds, despite the presence of dissatisfied agents.
\subsubsection{NN Search\label{subsubsec:NN(s=0)}}
For the NN search algorithm, at moderate occupation densities the first prospect sites' neighborhoods often share significant overlap with the agents' neighborhood and thus fail to produce a sufficient change in $q$ to meet the movement criteria. With only one prospect allowed, these agents who would otherwise initiate Schelling avalanches are immobilized, unable to relocate despite being unsatisfied. As a result, both movement criteria (m=1, m=2) yield petrified purist behavior for th $> 0.5$. These phase diagrams then ``collapse'' the class 2 petrified outcome to class 0 phase diagrams as the transitions overlap under constrained prospecting vision and time. With the upper transition being lowered to $\vartheta_m \simeq 0.5$ as a consequence of limited prospecting time and an inefficient search rule with high correlations between prospects. 

\subsubsection{Random Search\label{subsubsec:Random(s=2)}}
\begin{figure}
\includegraphics[width=8.7cm]{ 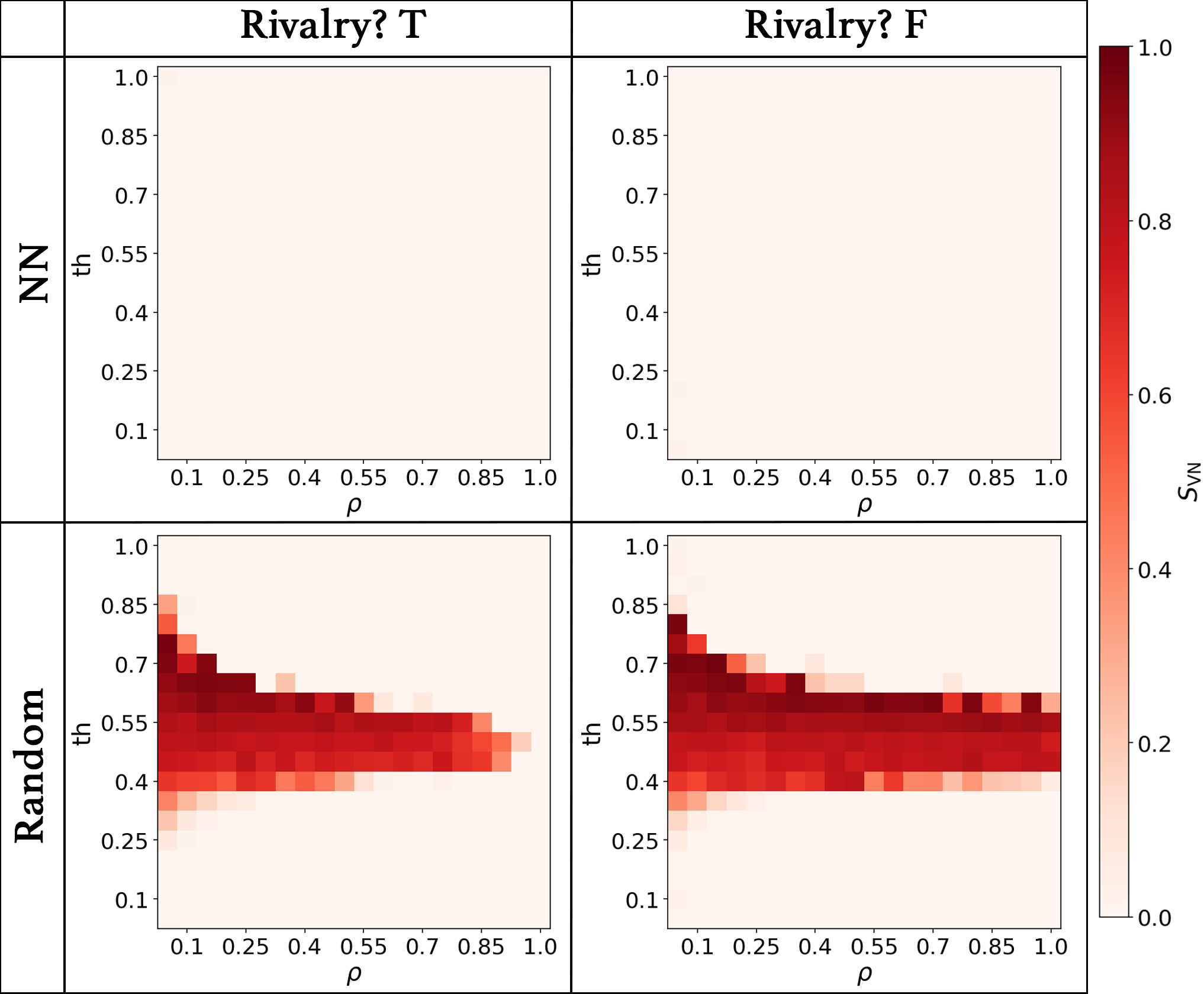}
\caption{\label{fig:prosp1-satisfied} Segregation outcomes of constrained prospecting. Segregation outcomes for satisfaction seeking agents when prospecting is constrained to 1 site. NN phase diagrams collapse to Class 0 as agents require $>1$ prospect due to inefficient search procedure. All simulations run with $\ell_p = 1$, $L=100$ and $r=10$.} 
\end{figure}
For random searches, when $\ell_p = L^2-1$ most avalanche seeds in the segregated region of the phase diagram relocated after prospecting one site. These randomly selected prospects typically share no spatial overlap with the agents original neighborhood, offering fresh opportunities to alter $q$. Thus, limiting prospected sites to 1 had little effect on the emergence of segregation at moderate thresholds across most occupation densities.

For improvement seeking agents (Class I), the number of sites prospected when th $> \vartheta_s$ depends on rivalry. When rivalry is enabled and the lattice is densely occupied $\rho > 0.85$, site availability decreases. The lack of site availability drives up the number of sites prospected resulting in immobilization when $\ell_p = 1$. This is evident when examining the right edge of the phase diagram (see Figure \ref{fig:prosp1-improved}). When rivalry is off, occupation density does not impact availability and most agents relocate within 1 search. 

For satisfaction seeking agents, the upper region of integration (th $> \vartheta_s$) corresponds to a lack of havens. In this regime, prospecting is futile as no satisfactory sites exist to relocate to. Consequently, constraining prospecting does not change the outcome in this region of the phase diagram.

\section{Control-Parameter Dependence of $\sigma_q$\label{sec:fluctuations}}
Starting with the first order Taylor expansion of q around its mean value $\langle q \rangle = 0.5$:
\begin{eqnarray}
q = \langle q \rangle  + a \cdot  \delta n_s + b \cdot \delta n_o
\label{eq:q-taylor}
\end{eqnarray} 
Where:
\begin{eqnarray}
a = \left(\frac{\partial q}{\partial n_s}\right)\bigg|_{\langle q \rangle}, \quad b = \left(\frac{\partial q}{\partial n_o}\right)\bigg|_{\langle q \rangle}
\end{eqnarray}
Where $\delta X$ denotes the fluctuation in a sample of X from its mean:
\begin{eqnarray}
\delta X = X-\langle X \rangle
\label{eq:fluctuation eqn}
\end{eqnarray} 
Substituting $q$ into the variance and applying properties of the variance, this simplifies to:


\begin{align}
& \operatorname{Var}\{q\} \nonumber \\
&\ \ = a^2\operatorname{Var}\{n_s\} + b^2 \operatorname{Var}\{n_o\}  + 2ab\operatorname{Cov}\{n_s, n_o\} 
\label{eq:Var qsame simp}
\end{align} 
\subsection{Expected neighborhood occupation}
In a neighborhood of $M = \pi r^2$ sites (in the large r limit)  the number of same agents is calculated as:
\begin{equation}
n_s = \sum_i^M J_i
\label{eq:Same count}
\end{equation} 
Where $J_i$ is a Bernoulli random variable:
\begin{equation}
J_i = 
\begin{cases}
    1, & \text{if site $i$ contains an agent of the same type,} \\
    0, & \text{otherwise.}\\
\end{cases}
\label{eq:bern-vals}
\end{equation} 
With probabilities:
\begin{equation}
p(J_i = 1) = \frac{\rho}{2}; \quad p(J_i = 0) = 1-\frac{\rho}{2}
\label{eq:bern-prob}
\end{equation}
The expected site occupation of same type agents is simply their density, which is $\rho/2$ for a symmetric population:
\begin{equation}
\langle J_i \rangle = \frac{\rho}{2}
\label{eq:bern-exp}
\end{equation} 
Thus:
\begin{equation}
\langle n_s \rangle =  \sum_i^M \langle J_i \rangle = \frac{M\rho}{2}
\label{eq:exp-same}
\end{equation} 
And similarly:
\begin{equation}
\langle n_o \rangle =  \sum_i^M \langle L_i \rangle = \frac{M\rho}{2}
\label{eq:exp-opp}
\end{equation} 
Where $L_i$ is the Bernoulli random variable for opposite agents, sharing the same probability due to the symmetric population density.

These can then be substituted into the partial derivatives of q to yield:
\begin{align}
a & = \frac{1}{\langle n_s \rangle  +  \langle n_o\rangle} \left(1-\frac{\langle n_s \rangle}{\langle n_s  \rangle +  \langle n_o\rangle} \right) \nonumber \\ &= \frac{1}{2M\rho} 
\label{eq:a}
\end{align} 
\begin{equation}
b = -\frac{\langle n_s \rangle}{(\langle n_s \rangle + \langle n_o \rangle)^2}  = -\frac{1}{2M\rho}
\label{eq:b}
\end{equation} 
\subsection{Variance and covariance in neighborhood occupation}
\begin{equation}
\operatorname{Var}\{ n_s \} = \sum_i^M \operatorname{Var}\{ J_i \} + \sum_i^M \sum_{k \neq i}^{M-1}\operatorname{Cov}\{ J_i, J_k \}
\label{eq:Var Same count fin}
\end{equation} 
The variance and covariance of the Bernoulli random variables are:
\begin{equation}
\operatorname{Var}\{ J_i \} = \frac{\rho}{2} \left(1-\frac{\rho}{2}\right)
\label{eq:Var Ji}
\end{equation}
\begin{equation}
\operatorname{Cov}\{ J_i, J_k \} = \langle J_iJ_k\rangle   -\langle J_i\rangle \langle J_k\rangle  
\label{eq:Cov JiJk}
\end{equation} 
Where:
\begin{eqnarray}
\langle J_iJ_k\rangle = \sum J_i J_k p(J_i, J_k)
\label{eq:JiJk}
\end{eqnarray} 
the probability of selecting two sites in a neighborhood and finding them both to be occupied is:

\begin{align}
& p(J_i, J_k)  = p(J_i)p(J_k|J_i) \nonumber \\ & p(J_i=1, J_k=1) = \frac{N}{2(L+1)^2} \times \frac{\frac{N}{2}-1}{(L+1)^2-1}
\label{eq:prob-JiJk}
\end{align}
Assuming $(L+1)^2 \gg 1$ and  $N/2 \gg 1$, $\langle J_iJ_k \rangle \to \langle J_i \rangle \langle J_k \rangle$ and the covariance vanishes:
\begin{equation}
\operatorname{Cov}\{ J_i, J_k \} \simeq  0
\label{eq:Cov JiJk fin}
\end{equation} 
Similarly for:
\begin{equation}
\operatorname{Cov}\{n_s, n_o\} \propto \operatorname{Cov}\{J_i, L_k\}
\label{eq:Cov n_sn_o}
\end{equation}
For $(L+1)^2 \gg 1$ and  $N \gg 1$, $\langle J_iL_k \rangle \to \langle J_i \rangle \langle L_k \rangle$ and the covariance vanishes:
\begin{equation}
\operatorname{Cov}\{n_s, n_o\} \propto \operatorname{Cov}\{J_i, L_k\} \simeq 0
\label{eq:Cov n_sn_o simp}
\end{equation}
Thus $\operatorname{Var}(n_s)$ and $\operatorname{Var}(n_o)$ dominate equation \ref{eq:Var qsame simp} with:
\begin{equation}
\operatorname{Var}\{ n_s \} \simeq  M\frac{\rho}{2} \left(1-\frac{\rho}{2}\right)
\label{eq:Var Same count}
\end{equation} 
It can easily be shown that the same result holds for $\operatorname{Var}\{n_o\}$ when the population is symmetric. 


\subsection{Putting it all back together}
Due to the symmetry of the population distribution and independent sampling of the population  \ref{eq:Var qsame simp} simplifies to:
\begin{equation}
\operatorname{Var}\{q\} = (a^2+b^2)\operatorname{Var}\{n_s\} 
\label{eq:Var qsame simp2}
\end{equation} 
Substituting in equations \ref{eq:a}, \ref{eq:b} and \ref{eq:Var Same count fin} where:
\begin{equation}
(a^2+b^2) = \frac{1}{2(M\rho)^2}
\end{equation} 
\begin{equation}
\operatorname{Var}\{ n_s \}\simeq  M\frac{\rho}{2} \left(1-\frac{\rho}{2}\right)
\end{equation}
This yields a variance that scales like by $\langle n \rangle^{-1}$, as seen in Figure \ref{fig:fluctuations-all}, with a coefficient that has weak dependence on $\rho$:
\begin{equation}
\operatorname{Var}\{q\} \simeq  \frac{1}{2\langle n \rangle} \left(1-\frac{\rho}{2}\right)
\label{eq:Var qsame final}
\end{equation} 

Where $\langle n \rangle = \langle n_s \rangle +\langle n_o \rangle $ is the expected number of occupants in a neighborhood.

\nocite{*}
\newpage
\bibliography{apssamp}

\end{document}